\documentclass[a4paper]{article}

\usepackage[dvips]{graphicx}
\DeclareGraphicsExtensions{.eps}
\graphicspath{ {figures/} }

\title{A Review of Procedures to\\Evolve Quantum Algorithms\\}
\author{\newline\textbf{Adrian Gepp and Phil Stocks}\\
School of Information Technology\\
Bond University\\
Queensland, Australia\\
}
\date{August 24, 2007}

\newcommand{\bra}[1]{\langle \hspace{0.01in} #1 \hspace{0.01in} |}
\newcommand{\ket}[1]{| \hspace{0.01in} #1 \hspace{0.01in} \rangle}
\newcommand{\innerp}[2]{\langle \hspace{0.01in} #1 \hspace{0.01in} | \hspace{0.01in} #2 \hspace{0.01in} \rangle}
\newcommand{\outerp}[2]{| \hspace{0.01in} #1 \hspace{0.01in} \rangle \langle \hspace{0.01in} #2 \hspace{0.01in} |}
\newcommand{\tensor}[2]{#1 \otimes #2}
\newcommand{\bell}[4]{\ket{\psi_{#1}} = \frac{\ket{#2} #3 \ket{#4}}{\sqrt{2}}}

\begin{document}
\maketitle

\abstract{There exist quantum algorithms that are more efficient than their classical counterparts; such algorithms were invented by Shor in 1994 and then Grover in 1996. A lack of invention since Grover's algorithm has been commonly attributed to the non-intuitive nature of quantum algorithms to the classically trained person. Thus, the idea of using computers to automatically generate quantum algorithms based on an evolutionary model emerged. A limitation of this approach is that quantum computers do not yet exist and quantum simulation on a classical machine has an exponential order overhead. Nevertheless, early research into evolving quantum algorithms has shown promise.\newline

This paper provides an introduction into quantum and evolutionary algorithms for the computer scientist not familiar with these fields. The exciting field of using evolutionary algorithms to evolve quantum algorithms is then reviewed.
}

\section{Introduction and Overview}
Quantum algorithms and evolutionary algorithms are two increasingly popular research fields. Although still relatively new areas of research, there is a large number of publications in both areas. The idea of using evolutionary algorithms to produce quantum algorithms, known as evolving quantum algorithms, has not been pursued to the same extent as research in the two fields independently. This new field of evolving quantum algorithms is in an interesting development stage with the potential to dramatically change the area of quantum computing.

This is a review paper in the field of \emph{evolving quantum algorithms using genetic programming}. The objectives of this paper are to
\begin{itemize}
	\item provide a detailed introduction to the field of quantum computing,
	\item introduce the concept of evolutionary algorithms, specifically genetic programming, and
	\item discuss the application of genetic programming to quantum algorithms, including a comprehensive summary of work in this new field.
\end{itemize}
These objectives will be addressed in order; this paper is directed at readers with general knowledge of computer science, but not necessarily of quantum computing or evolutionary algorithms.

\section{Introduction to Quantum Computing}
\label{sec:QC}
Since Feynman's famous speech delivered in December 1959 in which he mentioned the possibility that ``sub-microscopic'' computers could be built \cite{Feynman1959}, the world has seen great advances in computing power and computer miniaturization, especially in the last two decades\footnote{A more detailed and structured account of Feynman's idea was given in his book \cite{Feynman1961}.}. Nonetheless, there are limits to computer miniaturisation with current microchip technology \cite{PRS1998} and we still, as yet, have not realised Feynman's sub-microscopic prediction. The desire for ``sub-microscopic'' computers is fueled by the appeal of the ability to significantly increases the efficiency of storing, copying, transmitting and processing information using a computer that will not occupy significant physical space.

The two major approaches that have been proposed for sub-microscopic computing are \emph{quantum computing} and \emph{DNA computing}, which have been widely discussed theoretically and are the subject of a large number of theoretical and empirical studies. DNA or molecular computing, not to be confused with evolutionary or genetic algorithms, essentially involves using DNA molecules instead of microchips. This allows the information-processing capabilities of organic molecules to replace digital switching primitives and achieve this sub-microscopic size \cite{PRS1998}. As this section of this paper only introduces quantum computing, more information on DNA computing is contained in the excellent book by P\u{a}un, Rozenberg and Salomaa \cite{DNA1998} or the shorter article by Gro{\ss} \cite{Gross1998}.

	\subsection{Brief History of Quantum Computing}
	Quantum computing is derived from a link between quantum mechanics, computer science and classical information theory \cite{Leier2004}. Essentially quantum quantum computing can be viewed as developing algorithms that will run on quantum computers, which are based on quantum mechanics. In the 1920s, the theory of quantum mechanics was proposed: the well-known major contributors were Born, Dirac, Heisenberg, and Schr\"odinger \cite{Leier2004}. However, it wasn't until 1980 that the area of quantum computing truly began, when Benioff \cite{Ben1980} presented his \emph{quantum Turing machine}. It proved that quantum systems could coherently perform computation to solve problems.
	
	Two years later in 1982, Feynman \cite{Feynman1982} observed that quantum computing could not always be simulated efficiently on a classical computer, which lead to speculation that quantum computing could be more efficient than classical computing in some cases. In 1985, Deutsch \cite{Deutsch1985} further developed Benioff's quantum Turing machine and suggested that quantum computers might be able to solve problems that had no efficient solution on a classical or probabilistic Turing machine. He also presented an example oracle problem\footnote{An oracle problem is such that some properties of a function are trying to be discovered, whereby that function is given as a black box. The code of a black box function is unknown, but the function's results for given inputs can be calculated.}, now known as Deutsch's problem, that was solved more efficiently on a quantum computer than by any classical algorithm. In addition, four years later Deutsch \cite{Deutsch1989} developed another, more popular, model for quantum computing known as the \emph{quantum circuit model}. Nevertheless, neither Deutsch's problem nor its latter developed generalisation Deutsch-Jozsa's problem \cite{DeutschJozsa1992} proved to be practically useful, so the field continued to develop slowly.
	
	Peter Shor \cite{Shor1994} shocked the computing world in 1994 when he presented two efficient quantum algorithms, for factoring integers and the `discrete-log' problem, for which there was, and still is, no efficient classical counterpart. Researchers had been searching for an efficient factoring algorithm for over 20 years, and most people were confident that this algorithm did not exist as the most efficient classical algorithm discovered is the number field sieve \cite{Lenstra1993} that requires exponential time. In contrast, Shor's factoring algorithm requires only polynomial time. Shor's algorithm was also the first non-oracle quantum algorithm produced. Shor's algorithm was inspired by Simon's quadratic-time quantum algorithm \cite{Simon1994} that solves an oracle problem, which requires exponential time on a classical computer\footnote{Simon's algorithm, like Deutsch's, lacks practical application.}. Nevertheless, all the focus was on Shor's factoring algorithm, because of its extremely important applications in cryptography. All classical cryptography techniques, such as RSA, would be easily breakable with a quantum computer running Shor's algorithm. Therefore, it was this algorithm that spurred much interest and research in quantum computing.
	
	Even with this increase in research as a consequence of Shor's breakthrough, the only breakthrough since then has been Grover's search algorithm \cite{Grover1996} developed in 1996. Grover's algorithm searches an $n$-element unstructured list in $O(\sqrt{n})$, compared to the classical $O(n)$. Thus, Grover's algorithm only provides a quadratic speed-up, but it is a major breakthrough due to its wide-spread application in search problems. Since 1996, the only new quantum algorithms have been variants of Shor's and Grover's algorithms. Possible reasons for the slow discovery of quantum algorithms are presented in Section~\ref{sec:potent}.
	
	\subsection{Preliminaries}
	Depending on the reader's knowledge of linear algebra, computer science and quantum computing, some of the topics introduced in this section may already be familiar to the reader. Therefore, as these topics can be read independently, it is possible to just refer to those topics of personal interest.
	
		\subsubsection{Dirac Notation}
			Quantum states can be described as vectors, which are by convention expressed in the notation invented by Paul Dirac in 1958 \cite{Dirac1958}. The basic two elements of the Dirac notation are called bras and kets. Standard column vectors in a Hilbert Space\footnote{Chapter 16 of Hardy and Steeb's book \cite{HardySteeb2001} contains a good introduction to Hilbert Space.} are represented by kets, such as $\ket{\upsilon}$ for a column vector $\upsilon$. The matching bra is a row vector, denoted $\bra{\upsilon}$, which represents the conjugate transpose, or dual, of $\upsilon$. The conjugate transpose of $\upsilon$ is defined as the transpose of the vector, in which each element is the complex conjugate of the corresponding element in $\upsilon$. These single vector representations can then be combined to represent such operations as the inner and outer product of vectors; for example, the inner and outer product of $\upsilon$ and $\psi$ are written as $\innerp{\upsilon}{\psi}$ and $\outerp{\upsilon}{\psi}$ respectively.
		
		\subsubsection{Tensors and Tensor Products}
			In essence, a tensor is a geometrical identity that is the generalisation of $n$-dimensional vectors with their associated linear operators. Tensors can also be represented as multi-dimensional arrays. However, from a quantum computing standpoint, the tensor product is the most important operation that can be performed on tensors: the tensor product of $\upsilon$ and $\psi$ is denoted $\tensor{\upsilon}{\psi}$.
			
			The tensor product is the most general bilinear operation: it is a generalisation of the matrix product operation, whereby all values contained in an operand are multiplied independently with all values in the other operand. Thus, the result of $\tensor{\upsilon}{\psi}$ has the following properties:
			\begin{eqnarray*}
					Rank(\tensor{A}{B}) & = & Rank(A) + Rank(B)\\
					Dimension(\tensor{A}{B}) & = & Dimension(A) \times Dimension(B)
			\end{eqnarray*}
			The actual operation of the $\otimes$ operator is more clearly described by the following example:
			\begin{displaymath}
				A \otimes
				\left( \begin{array}{ccc}
					a & b \\
					c & d  \\
				\end{array} \right)
				=
				\left( \begin{array}{ccc}
					A\times a & A\times b \\
					A\times c & A\times d  \\
				\end{array} \right)
			\end{displaymath}
			A complete treatment of the mathematics of tensor products is unnecessary here considering the scope of this paper; for a more complete treatment of tensor products refer to Hungerford's book \cite{Hunger1974}. The application of tensor products in quantum computing is discussed in Section \ref{sec:qubits} and \ref{sec:entangle}.
			
			\subsubsection{Quantum Bit}
			\label{sec:qubit}
			A quantum bit, commonly referred to as a qubit, is the basic unit of information in quantum computing: it is analogous to the bit in classical computation. Similar to a classical bit, a qubit has two computational \emph{basis states}, sometimes referred to as \emph{eigenstates}, usually represented as $\ket{0}$ and $\ket{1}$, that correspond to a bit's 0 and 1 states respectively. However, unlike a classical bit, a qubit can be in a superposition of these two basis states. Due to this superpositional nature of a qubit it can be thought of as a 2-dimensional vector (in complex vector space) of length one with the above two basis states as orthonormal vectors:
		\begin{eqnarray*}
			\ket{0} & = & \left( \begin{array}{ccc} 1 \\ 0 \\ \end{array} \right) \\
			\ket{1} & = & \left( \begin{array}{ccc} 0 \\ 1 \\ \end{array} \right) \\
		\end{eqnarray*}
		Thus, a qubit in a superposition of basis states can be described as being a unit vector that lies between the two basis states. A qubit in an arbitrary state is therefore expressed as a linear combination of basis states:
		\begin{eqnarray*}
			& \alpha \ket{0} + \beta \ket{1} \hspace{0.25cm} or \hspace{0.25cm} \left( \begin{array}{ccc} \alpha \\ \beta \\ \end{array} \right), & \\
			& ~where~ \alpha,\beta ~are~complex~numbers~and~ | \alpha |^{2}+| \beta | ^{2}=1 &
		\end{eqnarray*}
		Nevertheless, just like a classical bit, a quantum bit can be read to get its value, which can only be one of the basis states. However, the outcome is not deterministic as in classical computing, but rather probabilistic. Given the above expression for a qubit, the probability of the qubit being measured in each basis state is determined by the values of $\alpha$ and $\beta$, which are referred to as \emph{amplitudes} or \emph{probability amplitudes}. The actual probability of the qubit being measured in a basis state is the square of the corresponding amplitude; for example, the probability of the qubit being measured as a $\ket{0}$ is $| \alpha | ^{2}$. Hence, the need for the condition $| \alpha |^{2}+| \beta | ^{2}=1$.
		
		As the two amplitudes of a quibit in a basis state are 1 and 0 (order irrevelant) it will have 100\% probability of being measured in the state it is in and 0\% probability of being measured in the alternative basis state; that is, a qubit in a basis state is measured deterministically. On the other hand, measurement of a qubit in a superposition of basis states changes the qubit into the basis state in which it is measured. The ramifications of this are that subsequent measurements of a qubit initially in superposition are deterministic and will yield the same outcome with 100\% probability. Therefore, a qubit in a basis state can be thought of as a classical resource; in contrast, a qubit in a superposition of basis states can be thought of as a purely quantum resource \cite{Tonder2004}.
		
		\subsubsection{Multiple Qubits}
		\label{sec:qubits}
		Individual bits in classical computation combine through cartesian product, but quantum bits combine through tensor product \cite{RieffelPolak2000}. Taking the simplest case of multiple qubits, two qubits, we can find the possible basis states as the tensor product of the two individual sets of basis states:
		\begin{eqnarray*}
			& \left\{\ket{0},\ket{1}\right\} \otimes \left\{\ket{0},\ket{1}\right\} & \\
			& = \left\{\ket{0}\otimes\ket{0},\ket{0}\otimes\ket{1},\ket{1}\otimes\ket{0},\ket{1}\otimes\ket{1}\right\} & ~or~more~concisely:\\
			& \left\{\ket{00},\ket{01},\ket{10},\ket{11}\right\} &
		\end{eqnarray*}
		The final concise version represents an extension of Dirac notation whereby $\ket{q_{0}q_{1}\ldots q_{n}}$ represents the basis for qubits $q_{0}\ldots q_{n}$. Thus, with $k$ written as a binary number, a two-qubit quantum system can be described as:
		\begin{displaymath}
			\sum_{k=0}^{3}{\delta_{k}\ket{k}} ~or~
			\left( \begin{array}{ccc} \delta_{0} \\ \delta_{1} \\ \delta_{2} \\ \delta_{3} \\ \end{array} \right) \\, ~where~
			\delta_{k} ~are~complex~numbers~and~ \sum_{k=0}^{3}{| \delta_{k} |^{2}} = 1
		\end{displaymath}
		Thus, an n-qubit quantum system can be written as $\sum_{k=0}^{2^{n}-1}{\delta_{k}\ket{k}}$, where $\sum_{k=0}^{2^{n}-1}{| \delta_{k} |^{2}} = 1$ and $| \delta_{k} |^{2}$ is the probability of the system being measured in the basis state $\ket{k}$. It is possible to measure a sub-set of the qubits in a multi-qubit system. It is also possible to determine such information as whether two qubits are equal without learning their value; however, these more complex measurements are equivalent to a transformation of the quantum system followed by a standard measurement to determine the basis state of each qubit \cite{RieffelPolak2000}, and consequently it is common practise to also refer to standard measurements.
	
	\subsubsection{Entangled States}
	\label{sec:entangle}
	A suprising and non-intuitive aspect of quantum computing is that there are quantum states that can not be described in terms of their individual component qubits. These states are known as \emph{entangled states}, and the individual qubits within entangled states are known as \emph{entangled particles}. The reason for the use of the term `entangled' should become obvious in the next paragraph.
	
	We established in the previous section that single qubits combine using the tensor product operator. Therefore, two qubits in superposition combine as follows:
	\begin{eqnarray*}
		\ket{q_{1}} \otimes \ket{q_{2}} & = & \left(\alpha_{1} \ket{0} + \beta_{1} \ket{1}\right) \otimes \left(\alpha_{2} \ket{0} + \beta_{2} \ket{1}\right) \\
		& = & \alpha_{1}\alpha_{2}\ket{00} + \alpha_{1}\beta_{2}\ket{01} + \alpha_{2}\beta_{1}\ket{10} + \beta_{1}\beta_{2}\ket{11}
	\end{eqnarray*}
	However, there are entangled states that can not be described by the above formula. The canonical examples are the Bell states:
	\begin{displaymath}
		\begin{array}{cc}
			\bell{00}{00}{+}{11} & \bell{01}{01}{+}{10} \\
			\bell{10}{00}{-}{11} & \bell{11}{01}{-}{10} \\
		\end{array}
	\end{displaymath}
	In order to write $\psi_{00}$ in terms of the above two-qubit equation $\alpha_{1}\beta_{2}$ must equal $0$. This subsequently implies that either $\alpha_{1}\alpha_{2}=0$ or $\beta_{1}\beta_{2}=0$; however, it is clear that neither of these terms are zero. Thus, $\psi_{00}$ is an entangled state (the other Bell states are shown to be entangled with similar reasoning). An example of how a non-entangled state can be expressed in terms of its component qubits is the state $\frac{1}{sqrt{2}}\left(\ket{10} + \ket{11}\right) = \ket{1} \otimes \frac{1}{sqrt{2}}\left(\ket{0} + \ket{1}\right)$
	
	Another way of viewing entangled states is that measurement of one qubit has an effect on the other qubits. Take $\psi_{00}$ as an example again, if the first qubit is measured as $\ket{0}$ then we know that the second qubit will also be measured as $\ket{0}$. In contrast, measurement of the first and second qubit in our non-entangled state will always yield $\ket{0}$ for the first and a fifty-fifty chance of either basis state for the second, regardless of the measurement of the measurement order. Some famous research explaining this observation includes that of Einstein, Podolsky and Rosen: an accurate and concise summary of this work is given by Rieffel and Polak \cite{RieffelPolak2000}. It is clear that there is no classical counterpart of entangled states. However in quantum computing, it is important to realise that entangled states are treated no differently from any other quantum states. Nevertheless, they are extremely important and some very interesting applications of entangled state have been found such as \emph{dense coding}, \emph{quantum teleportation} (see Section~\ref{sec:teleport}) and secure \emph{quantum key distribution}, which are discussed by Rieffel and Polak \cite{RieffelPolak2000}. A detailed analysis of entangled states, their importance and a C++ implementation, is provided by Hardy and Steeb \cite{HardySteeb2000}.
			
	\subsection{Quantum Computers}
	\label{sec:QCrealise}
	Quantum computers still remain abstract machines despite the large amount of money and time that has been invested into building them. Quantum computers with a few qubits have been developed; however, a scalable general-purpose quantum computer is yet to be built. The largest quantum computer developed thus far consists of seven qubits \cite{SClark2005}, which was built by a group of scientists from Stanford University. It was constructed using Nuclear Magnetic Resonance (NMR) and tested using Shor's factoring algorithm \cite{Nature2001}. In February 2007, the Canadian company D-Wave publicly demonstrated `Orion', which they claim is a 16-qubit quantum computer. However, details of Orion's inner workings are unknown as no academic papers have been published about Orion\footnote{One of many news articles about Orion can be viewed at\newline http://arstechnica.com/articles/paedia/hardware/quantum.ars.}. Thus, there is academic speculation about whether Orion is truly a \emph{quantum} computer. There have been many other approaches to building a quantum computer, including optical photon spins, quantum dots, cavity quantum electrodynamics and ion traps. NMR and ion traps are the most advanced approaches to date; however, all of these techniques have had scaling problems. Therefore, it is commonly argued whether or not quantum computers of a significant scale (greater than 100 qubits) will ever be practically realised. Pellizzari \cite{Pellizzari1998} presents a general overview of the requirement for building a quantum computer, including its challenges, various approaches and future outlook; however, this article was published in 1998 and consequently does not include the new NMR approach. Hardy and Steeb \cite{HardySteeb2001} provide a concise summary of NMR, and Nielsen and Chuang \cite{Nielsen2000} provide a more detailed look at all of the above mentioned approaches.
	
	There are certain characteristics that an operational quantum computer must have regardless of the approach adopted for its construction. Different publications outline various characteristics that are somewhat similar \cite{HardySteeb2001,Loss1998,Nielsen2000}. These characteristics have been combined and summarised below:
	\begin{description}
		\item[Qubit Representation] There must be a representation of a qubit (dynamic two-state object) such as the up and down spins of a proton.
		\item[State Preparation] The system must reliably start in a known initial state. It is sufficient and common to start with all qubits in the same state (conventionally $\ket{0}$); thus, this requirement has not been a major issue.
		\item[State Transformation] There must be a mechanism to efficiently transform a quantum system into other states following the rules of quantum mechanics. The most common way this has been implemented is by the use of \emph{quantum gates}, which will be discussed in Section~\ref{sec:qgates}.
		\item[Measurement] There must be an efficient and reliable way to measure qubits in the system.
		\item[System Isolation] The system must be isolated from the environment to prevent the superposition of states from decaying quickly in the common environment, which is a phenomenon known as decoherence or quantum noise. This is a problem as algorithms may not be able to be applied in time if the states decay too quickly. This issue of decoherence is a major hurdle for the actualisation of a quantum computer \cite{RieffelPolak2000,Landry2004,Shor2000}. Zurek \cite{Zurek1991} provides a comprehensive and simple discussion of decoherence.
		
		It is almost inevitable that complete isolation will not occur and there will be some decoherence. As a partial solution, research is being undertaken on quantum error-correcting codes (QECCs). Since the first 9-qubit QECC was presented by Shor in 1995 \cite{Shor1995} research and development in this area has been making noticeable progress; the most recent work presents the most time-efficient QECC (although not the most space-efficient) \cite{ForneyGuha2005}. Despite progress being made in developing quantum error-correcting code, this method will always come at a cost of extra qubits, which consequently increases the number of qubits needed to implement any given algorithm. This is the reason QECC is solution for only a small amount of decoherence. Nielsen and Chuang \cite{Nielsen2000} present a comprehensive section on the area of quantum error-correction, which has been summarised by Landry \cite{Landry2004}. Furthermore, Hardy and Steeb \cite{HardySteeb2001} discuss the latest algorithms for developing quantum error-correcting code. 
	\end{description}
	
	Whether a scalable quantum computer will ever be built obviously has huge impacts for the field of quantum computing, as well as classical computing. Nevertheless, quantum computing is currently a popular research field and this paper simply views a quantum computer as an abstract machine. Furthermore, while this paper does not discuss quantum mechanics directly, most of the basic postulates of quantum mechanics will be discussed within Section~\ref{sec:QC} of this paper. For a structured coverage of the exact postulates of quantum mechanics refer to Nielsen and Chuang's book \cite{Nielsen2000}.	
	
	\subsection{Power of Quantum Computing}
	
	After more than 50 years of using the classical physics paradigm (which the classical Turing machine embodies) to build a theory of (standard) computation, quantum physics provides a different paradigm that is arguably more powerful than standard computation \cite{Lots1995}. The possibility of harnessing the power of quantum parallelism (discussed in the following section) and the identification of entangled states that are without a classical counterpart (discussed in Section~\ref{sec:entangle}) were the first indications that quantum computing could allow faster information processing than classical computing. Furthermore in 1985, Deutsch \cite{Deutsch1985} proposed that his quantum Turing machine could not only simulate quantum systems better than classical methods, but it may also be able to solve classical problems significantly faster than classical, and possibly probabilistic, Turing machines. This essentially suggested a violation in the strong Church-Turing thesis that any algorithmic process can be simulated on a classical Turing machine (CTM) in at worst a polynomial slowdown.
	
	\subsubsection{Quantum Parallelism}
	\label{sec:parallel}
	Quantum parallelism is the term used to describe the potential parallel computing power of quantum computing. It is intuitive that an increase in time efficiency can be gained by using parallel processors. Furthermore, in classical computation, an exponential time efficiency increase requires an exponential increase in the number of processors or physical space. However, as qubits can represent a superposition of two different states, in quantum computation, a linear increase in physical space generates an exponential increase in parallelism, and consequently an exponential increase in time efficiency. This is what is known as \emph{quantum parallelism} \cite{DeutschJozsa1992}. As shown in Section~\ref{sec:qubit}, a qubit can be represented as a two-dimensional vector. In classical physics, $n$ two-dimensional vector objects form a $2n$-dimension vector space. However due to quantum parallelism this is not the case in quantum computing in which $n$ qubits form a $2^{n}$-dimension vector space. This exponential growth implies a possible exponential increase in the information processing speed of quantum computers over classical computers.
	The concept of how quantum parallelism is applied to create parallel computation is conveniently thought of as shown below in a two-qubit example:
	\begin{eqnarray*}
		\begin{array}{ccccc}
			\left( \begin{array}{ccc} \delta_{0} \\ \delta_{1} \\ \delta_{2} \\ \delta_{3} \\ \end{array} \right) & \otimes & T & = & \left( \begin{array}{ccc} \delta_{0}\otimes T \\ \delta_{1}\otimes T \\ \delta_{2} \otimes T\\ \delta_{3} \otimes T\\ \end{array} \right) \\
		\mbox{All Inputs} & \otimes &	\mbox{One Transformation} & = & \mbox{All Outputs}
		\end{array}
	\end{eqnarray*}
	A set of input qubits in a superposition of basis states essentially has all the possible inputs encoded in them. Therefore, a \emph{single} transformation can be applied to generate a set of output qubits in a superposition of basis states, which represents all possible outputs. Thus, all output states have been computed and assigned a probability of measurement on a quantum computer in the same time it takes to compute the output for a single input state on a classical computer.
	
	Although all outputs can be produced using only one transformation, we know from Sections \ref{sec:qubit} and \ref{sec:qubits} that only one output can be extracted upon measurement. Furthermore, the particular output that will be extracted is unknown as measurement is probabilistic. Thus, a quantum algorithm must manipulate quantum parallelism so that the desired result is extracted when measured. This is a difficult and non-intuitive task that no classical programming technique can solve. The two best--known approaches are \cite{RieffelPolak2000}
\begin{itemize}
	\item Measure common properties of all the outputs. This approach is used in Shor's factoring algorithm whereby the period of the outputs is measured
	\item Increase the amplitudes of the basis states of interest to increase the probability that they will be measured. This technique is used in Grover's search algorithm.
\end{itemize}
	In addition, there are other restrictions on quantum computing that are discussed in the next section.
	
	\subsubsection{Unitary Restriction}
	\label{sec:QChurdle}
	With the exception of measurement, all transformations in a quantum system must be unitary. This is specified in Schr\"odinger's equation, which governs the dynamics of a quantum system which is not being measured. To be precise, Schr\"odinger's equation enforces that quantum systems must change state in a way that preserves orthogonality\footnote{To be orthogonal means to be linearly independent: a precise treatment of the orthogonality condition is given by MathWorld [available at http://mathworld.wolfram.com/OrthogonalityCondition.html].}, and the only transformations that maintain orthogonality in a complex vector space, such as quantum systems, are unitary transformations. The transformation matrix $M$ is a unitary matrix if $M M^{*} = I$, where $M^{*}$ is the conjugate transpose of M. A unitary transformation can be thought of as a rotation in complex vector space \cite{RieffelPolak2000}: this makes sense as vector rotations will maintain the angles between all vectors and preserve orthogonality.
	
	An important consequence of quantum transformations being unitary is that they must be reversible, as unitary matrices are invertible. For computing to be reversible, the current state must uniquely determine the previous state, for all computational states \cite{Bennett1973}. However, this is not a problem as Bennett \cite{Bennett1973} showed that all classical computation can be transformed into reversible computation without a polynomial expenditure in time or space. Another important consequence of unitary transformations is that their linearity implies that quantum states can not be cloned, which is known as the \emph{no-cloning theorem}. An easily understandable version of the original proof of this theorem by Wootters and Zurek \cite{Woot1982} is given by Rieffel and Polak \cite{RieffelPolak2000}. It is important however to note that this theorem only applies to qubits in an unknown state: qubits in a basis state or a known state of superposition\footnote{A known state of superposition is where all the amplitudes are known.} can be cloned, but qubits in an unknown state can not be cloned as the information about their amplitudes will be lost.
	
	\subsubsection{The Potential}
	\label{sec:potent}
	There is a well-established view that quantum computing will only yield an exponential speedup in problems whose structure avoids the need to process exponentially many cases \cite{Braun1995}. That is, a brute force approach to NP-complete problems will not succeed with the aid of quantum parallelism; Fortnow and Rogers \cite{FortnowRogers1999} also firmly question whether quantum computing is more powerful than classical computing, but they still maintain it is a potentially powerful model of computation. However, whether quantum computation can efficiently solve NP-complete problems in polynomial time remains an open question \cite{RieffelPolak2000}. With the discovery of a few quantum algorithms there is no doubt that certain problems can be solved more efficiently on a quantum computer than on a classical computer. The number and diversity of problems that can be solved more efficiently on a quantum computer is still unknown and the subject of much current research.
	
	In reality, it is not extremely difficult to develop a working quantum algorithm. However, quantum algorithms are only of interest if they are \emph{more} efficient than their classical counterparts. This is one of the main reasons that more quantum algorithms have not been discovered. It is simple enough to simulate a classical algorithm on a quantum computer; however, producing a quantum algorithm that is superior to its classical counterpart requires the use of truly quantum effects\footnote{The term quantum effects is used as an umbrella term to encompass effects such as superposition and entanglement that exist in quantum computing, but not classical computing.}, which is extremely complex. Pessimists speculate that the lack of discovery of quantum algorithms is due to a lack of application of quantum computing. However, it seems more likely the reasons are the non-intuitive nature of creating quantum algorithms to the classically trained world and the fact that only superior quantum algorithms are of interest. Nevertheless, it has not been conclusively proven that quantum computation is more powerful than classical computation (this issue will be discussed more in the following section). The potential of quantum computation is however extremely high as there may even be an exponential speed gap between a quantum and classical computer \cite{Gramss1998}, as is the case for Shor's factoring algorithm \cite{Shor1994}.
	
	\subsection{Complexity of Quantum Algorithms}
		Quantum computing is a unique type of probabilistic computing, which has caused the creation of new complexity classes \cite{RieffelPolak2000}. The most interesting quantum complexity class is bounded-error quantum polynomial time ($BQP$), which is the quantum equivalent of the classical bounded-error probabilistic polynomial time ($BPP$) complexity class. $BPP$ contains all languages for which there is a probabilistic Turing machine such that it gives the right answer with bounded probability, which means greater than or equal to $\frac{2}{3}$ of the time. Note that the probability $\frac{2}{3}$ is simply a norm as it can be replaced with any number between $\frac{1}{2}$ and $1$ without altering the complexity class: if the probability is set higher than $\frac{1}{2}$ any previously $BPP$ algorithm can simply be rerun to amplify the probability \cite{Shor2000}. Using this definition of $BPP$ and substituting `quantum Turing machine' for `probabilistic Turing machine' we obtain the definition of the complexity class $BQP$.
		
		It is a significant result in quantum complexity theory that Bernstein and Vazirani proved that $BPP \subseteq BQP \subseteq PSPACE$ \cite{BV1993}, but the open question of extreme importance in quantum computing is whether $BQP$ is strictly larger than $BPP$. It is important to note that proving $BPP \subset BQP$ implies that $BPP \subset PSPACE$, which is not currently known. Thus, proving that quantum computers are more powerful than classical computers would also represent a breakthrough in classical computing complexity theory. However, it also indicates that proving $BPP \subset BQP$ is probably extremely difficult. Simon's quadratic-time quantum algorithm \cite{Simon1994} to solve a problem with known classical complexity of exponential-time suggests that $BQP$ is strictly larger than $BPP$, but it is not a rigorous proof as it only shows $BPP \subset BQP$ for oracle problems \cite{Shor2000}. Grover's search algorithm \cite{Grover1996} also puts forward a case that $BQP$ is strictly larger than $BPP$; however, it remains an open question \cite{Leier2004}. That is, it has not yet been proven that quantum computers have more capabilities than classical computers.
		
		Further details on quantum complexity classes are outside the scope of this paper, but more information on the $BQP$ complexity class is presented by Fortnow and Rogers \cite{FortnowRogers1999}, and \cite{Bennett1997,Watrous1998} detail quantum computing complexity in time and space respectively.
	
	\subsection{Models of Quantum Computation}
	 The first quantum computation model was the quantum Turing machine (QTM) introduced by Benioff \cite{Ben1980} and Deutsch \cite{Deutsch1985}, which can efficiently simulate every classical Turing machine (CTM) with a polynomial-bounded overhead \cite{Gramss1998}. The QTM is also the commonly used basis for defining universality. In essence, the QTM has three deterministic CTM tapes where the two extra tapes are needed to make the computation reversible (the set of tape states also changes to reflect the nature of quantum computing). The QTM is described in more detail by Gram\ss~\cite{Gramss1998}.
	 
	 There are also many other mathematically equivalent models of quantum computation, just as there are many different models of classical computation. Moreover, there are other potential models of quantum computation that have not been explored \cite{Shor2000}. A quantum state machine \cite{Gudder1999} has been proposed that is an extension of classical finite state machines in which amplitudes are added to transitions to represent quantum parallelism; a universal quantum cellular automata model \cite{vanDam1996} is another quantum computing model. Recently, a universal quantum lambda calculus \cite{Tonder2004} has also been suggested that is based upon the classical lambda calculus. Essentially, a reversible lamdba calculus is established with the use of minor reduction steps, and the quantum parallelism is captured as subexpressions representing superposition that can not be reduced/observed (no-cloning theorem).
	 
	 This section of this paper focuses on the quantum circuit model, due to its popularity and simplicity. The quantum circuit model, also known as the quantum gate array, was introduced by Deutsch in 1989 \cite{Deutsch1989}. The circuit model is mathematically equivalent to the QTM \cite{Yao1993}, and is the most popular model of quantum computation. The reason for the popularity difference is that the QTM suffers from the same complexities as the CTM; it does not satisfy a simple algebra and it can be cumbersome to use as it requires `word at a time' thinking while keeping track of control variables such as tape states \cite{Tonder2004}. In contrast, quantum circuits are easier to understand as they are graphical diagrams, which can also be manipulated algebraically. Furthermore, other models such as the quantum lambda calculus are too new to be widely popular. The quantum circuit model will be explained in more detail in the following sub-section.

		\subsubsection{Quantum Circuit Model and Quantum Gates}
		\label{sec:qcircuit}
		Similar to classical circuits, the quantum circuit model consists of wires, and gates that act upon wires. In the quantum circuit model, each wire represents a qubit\footnote{By convention a qubit begins in a basis state, usually assumed to be $\ket{0}$.} and each individual gate has the same number of input qubits as it has output qubits (due to the reversibility of quantum computing). Thus, an advantage of this model is that the diagram itself can be essentially thought of classically, with the data on the wires representing the quantum nature of the computing. As the number of qubits (wires) is required to be constant, an important restriction of this model is that a quantum circuit can only compute a function with a specific length domain. Therefore, for functions with arbitrary length inputs a \emph{family} of quantum circuits is required; that is, a quantum circuit for each input length is required.

		The complexity of a quantum circuit depends on the total number of gates and qubits (wires). However, as gates are commonly restricted to act upon one, two or three qubits\footnote{There is some inconsistency in the literature about the maximum number of qubits that a gate should act upon; for example, Shor \cite{Shor2000} states the maximum should be two qubits while Spector et al. \cite{Spector1998} state that the maximum is a few qubits.}, the number of gates in a quantum circuit is both a reasonable and the usual measure of complexity. This restriction does not affect the universality of quantum circuits; for example, the set of all one-qubit gates and the CNOT gate (explained later in this section) are universal for quantum computing. Further note that an interesting three-qubit gate is the Toffoli gate that is in fact universal for quantum computing as presented by Deutsch \cite{Deutsch1989}. In addition, it has more recently been shown that a large number of two-bit quantum gates are universal \cite{Barenco1995}. An excellent review of quantum gates and the respective universality issue is contained in \cite{Lots1995}.
		
		\label{sec:qgates}
		Quantum gates are in fact unitary transformation matrices (see Section~\ref{sec:QChurdle}), and a matrix that acts upon $n$-qubits will be a $2^{n} \times 2^{n}$ square matrix due to quantum parallelism. Quantum parallelism (see Section~\ref{sec:parallel}) is the reason for a linear increase in the number of qubits causing an exponential increase in the size of a quantum system. To familiarise the reader with both the gates and the quantum circuit model\footnote{Variations in the quantum circuit notation of different gates exist, but they are only minor and should be understood by a reader who has read this paper.}, some important elementary gates are detailed below using the quantum circuit model.

		Three important single-qubit gates are the identity (I), negation (X), and Hadamard (H) transformations, which are detailed below\footnote{Due to linearity, transformations are fully specified by their effect on the basis states \cite{RieffelPolak2000}.}:
		\begin{eqnarray*}
			I: &	\left( \begin{array}{ccc} 1 & 0 \\ 0 & 1 \\ \end{array} \right) & \ket{0}~\rightarrow~\ket{0}; \ket{1}~\rightarrow~\ket{1} \\
			X: &	\left( \begin{array}{ccc} 0 & 1 \\ 1 & 0 \\ \end{array} \right) & \ket{0}~\rightarrow~\ket{1}; \ket{1}~\rightarrow~\ket{0} \\
			H: &	\frac{1}{\sqrt{2}}\left( \begin{array}{ccc} 1 & 1 \\ 1 & -1 \\ \end{array} \right) & \ket{0}~\rightarrow~\frac{1}{\sqrt{2}}\left(\ket{0}+\ket{1}\right); \ket{1}\rightarrow~\frac{1}{\sqrt{2}}\left(\ket{0}-\ket{1}\right)
		\end{eqnarray*}
		It is important to notice that the Hadamard gate creates an equal superposition of states, and when applied to n-qubits it creates a superposition of all $2^{n}$ states (this is the reason for its prevalence in quantum algorithms). The workings of these gates can be easily verified with linear algebra; for example, the negation transformation on $\ket{0}$:
		\begin{displaymath}
			 \left( \begin{array}{ccc} 0 & 1 \\ 1 & 0 \\ \end{array} \right) \left( \begin{array}{ccc} 1 \\ 0 \\ \end{array} \right) = \left( \begin{array}{ccc} 1\times0 + 0\times1 \\ 1\times1+ 0\times0 \\ \end{array} \right) = \left( \begin{array}{ccc} 0 \\ 1 \\ \end{array} \right)
		\end{displaymath}
		These single-qubit transformations are represented in the quantum circuit model by an appropriately labeled box as shown below:
		\begin{figure}[h]
\centerline{
   \includegraphics[scale=0.6]{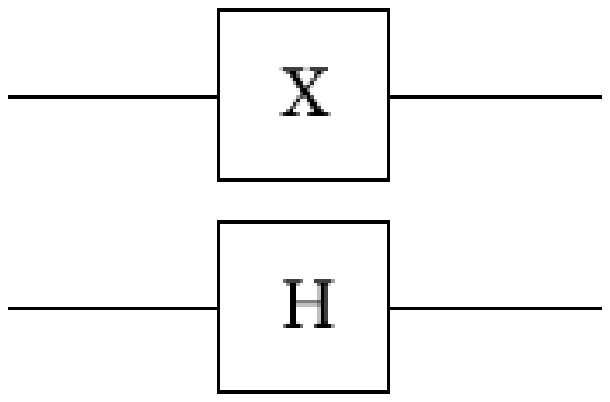}
}
\end{figure}
		
		The controlled-not (CNOT) is an extremely important two-qubit gate. Its importance comes from the fact that the results of a CNOT transformation can not be expressed as the tensor product of two qubits; thus, a CNOT gate can be used to create entangled states. The workings of the CNOT gate are relatively simple: it swaps the target qubit (represented by a cross) if the control qubit (represented by an open circle) is $\ket{1}$. The mapping and quantum circuit notation for the CNOT gate are as shown.
		\begin{eqnarray*}
			& \ket{00}~\rightarrow~\ket{00} & \ket{01}~\rightarrow~\ket{01}\\
			& \ket{10}~\rightarrow~\ket{11} & \ket{11}~\rightarrow~\ket{10}
		\end{eqnarray*}
		\begin{figure}[ht]
\centerline{
   \includegraphics[scale=0.5]{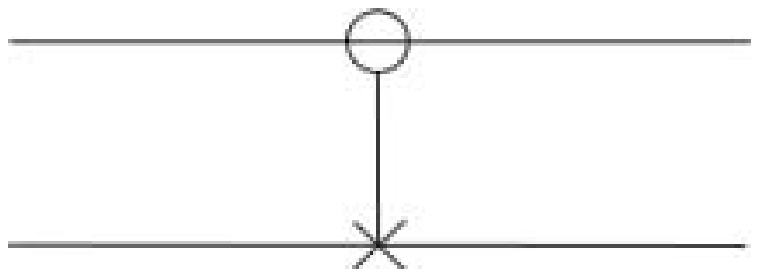}
}
\end{figure}
		
		The universal 3-qubit Toffoli gate is essentially a controlled-controlled-not gate, where the target qubit is only flipped if the two control qubits are $\ket{1}$; It is represented in the circuit model as follows.
		\begin{figure}[ht]
\centerline{
   \includegraphics[scale=0.5]{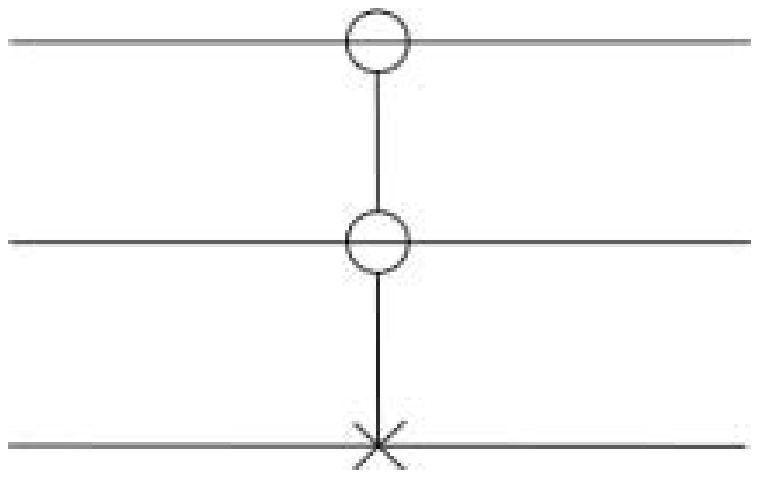}
}
\end{figure}
	
	\subsection{Example Problems and Algorithms}
	The algorithms that will be presented in this section are quantum teleportation, and an overview of Grover's algorithm. These algorithms have been chosen for their ease of understanding and their ability to convey important quantum algorithm concepts. Pittenger's book \cite{Purple} contains excellent, succinct treatments of the following quantum algorithms: Deutsch-Joza's, Simon's, Grover's, Shor's and the Quantum Fourier Transformation.
	
	This section will also include definitions of Deutsch's and the scaling majority-on problems. Quantum algorithms to solve these problems have been evolved by Spector et al. using genetic programming (see Section~\ref{sec:spector}).
	
	\subsubsection{Quantum Teleportation Algorithm}
	\label{sec:teleport}
	Quantum teleportation\footnote{This explanation of the quantum teleportation algorithm is based on the explanation given by Landry \cite{Landry2004}.}, uses two classical bits to transfer one quantum bit of information ($\ket{q1}$) from A(lice) to B(ob) without ever being anywhere in between\footnote{Quantum teleportation of one qubit has been realised experimentally \cite{Bouw1997}.}. This does not violate the no-cloning theorem as the unknown qubit is teleported not copied; that is, after the quantum teleportation the qubit no longer exists with Alice. The quantum circuit model for teleportation is as shown below in Figure~\ref{fig:tporta}.
	\begin{figure}[ht]
\centerline{
   \includegraphics[scale=0.5]{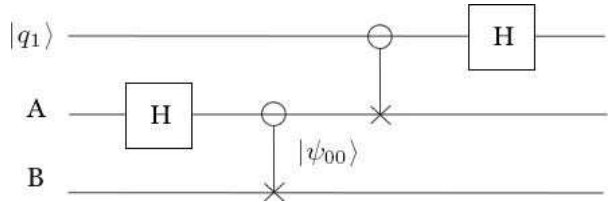}
}
\caption{Quantum Teleportation Algorithm}
\label{fig:tporta}
\end{figure}

	The first step creates Bell's first entangled state $\bell{00}{00}{+}{11}$ (discussed in Section~\ref{sec:entangle}) by putting the first qubit into equal superposition using the Hadamard gate and then modifying the second according to the value of the first using the CNOT gate. Now, given that $\ket{q1}$ is unknown, it can be represented as $\alpha\ket{0}+\beta\ket{1}$, and the quantum state after the first two transformations is
	\begin{displaymath}
		\frac{1}{\sqrt{2}}\left(\alpha\ket{0}\left(\ket{00}+\ket{11}\right)+\beta\ket{1}\left(\ket{00}+\ket{11}\right)\right)
	\end{displaymath}
	Alice and Bob are each given one of these two entangled particles and the final two transformations are performed as shown, and the next CNOT is applied to make the quantum state
	\begin{displaymath}
		\frac{1}{\sqrt{2}}\left(\alpha\ket{0}\left(\ket{00}+\ket{11}\right)+\beta\ket{1}\left(\ket{10}+\ket{01}\right)\right)
	\end{displaymath}
	Thus, the information about $\ket{q1}$ is now contained in the entangled pair, and the final Hadamard transformation is performed to yield
	\begin{eqnarray*}
&\frac{1}{2}\left(\alpha\left(\ket{0}+\ket{1}\right)\left(\ket{00}+\ket{11}\right)+\beta\left(\ket{0}-\ket{1}\right)\left(\ket{10}+\ket{01}\right)\right)& \\
&=\frac{1}{2}\left[\ket{00}\left(\alpha\ket{0}+\beta\ket{1}\right)+	\ket{01}\left(\alpha\ket{1}+\beta\ket{0}\right) +  \ket{10}\left(\alpha\ket{0}-\beta\ket{1}\right) + \ket{11}\left(\alpha\ket{1}-\beta\ket{0}\right)	\right] & \\
	\end{eqnarray*}
	Alice then measures the two qubits she has and sends the result encoded in two classical bits \{00,01,10,11\} to Bob. Using the previous equation, Bob knows the state of his qubit in terms of $\ket{q1}$. Therefore, he can apply a simple one-qubit gate (if needed) to convert his qubit to $\ket{q1}$ as defined below (using the previous equation):
	\begin{eqnarray*}
	\mbox{Alice's Result}	& \mbox{Bob's Qubit}	& \mbox{Transformation Required} \\
	\ket{00} & \alpha\ket{0}+\beta\ket{1} & none \\
	\ket{01} & \alpha\ket{1}+\beta\ket{0} & negation (X) \\
	\ket{10} & \alpha\ket{0}-\beta\ket{1} & \left(\begin{array}{ccc} 0 & 1 \\ -1 & 0 \\ \end{array} \right) \\
	\ket{11} & \alpha\ket{1}-\beta\ket{0} & \left(\begin{array}{ccc} 1 & 0 \\ 0 & -1 \\ \end{array} \right)
	\end{eqnarray*}
	For example, in the case of $\ket{01}$, after transformation Bob's qubit equals
	\begin{eqnarray*}
		X\left(\alpha\ket{1}+\beta\ket{0}\right) & = & \left( \begin{array}{ccc} 0 & 1 \\ 1 & 0 \\ \end{array} \right) \left( \begin{array}{ccc} b \\ a \\ \end{array} \right) \\
		& = & \alpha\ket{0}+\beta\ket{1} \\
		& \equiv & \ket{q1}
	\end{eqnarray*}
	Thus, an unknown qubit in a superposition of states can be transported from A to B with only two classical bits of information.
		
	\subsubsection{Grover's Algorithm}
	Grover's algorithm searches an unstructured $n$-element list in $O(\sqrt{n})$ time compared to its classical counterpart which is $O(n)$. An overview of the algorithm in the quantum circuit model is shown below in Figure~\ref{fig:goutline} (based on Figure 6.1 in Nielsen and Chuang's book \cite{Nielsen2000}).
	
	\begin{figure}[h]
\centerline{
   \includegraphics[scale=0.5]{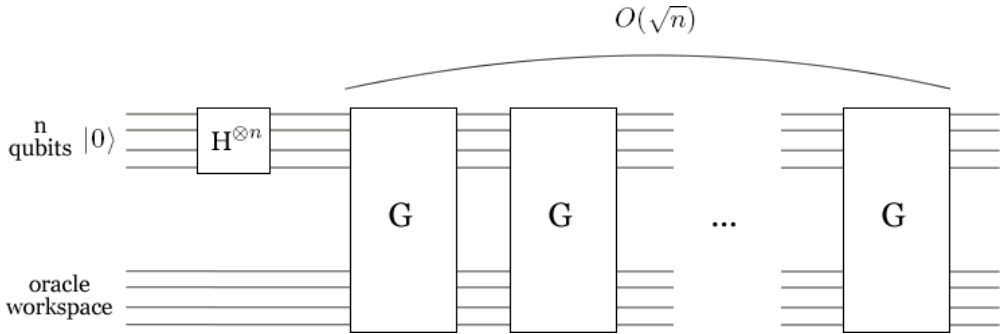}
}
\caption{Grover's Algorithm}
\label{fig:goutline}
\end{figure}

\begin{figure}[htb]
\centerline{
   \includegraphics[scale=0.23]{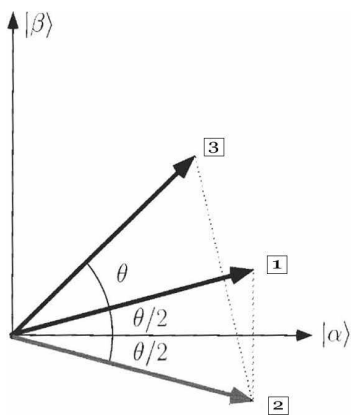}
}
\caption{The result from applying the Grover operator: modified from Figure 6.3 in Nielsen and Chuang's book \cite{Nielsen2000}. The three numbers in boxes indicate the initial ($1$), intermediate ($2$), and final ($3$) state vectors when the Grover operator is applied. $\ket{\beta}$ is the superposition of all solutions to the search problem, and $\ket{\alpha}$ is a state orthogonal to $\ket{\beta}$. All vectors in this diagram are unit vectors, but $\ket{\alpha}$ and $\ket{\beta}$ are shown to be longer to improve clarity. }
\label{fig:gop}
\end{figure}

	The gate labeled $H^{\otimes n}$ is a shorthand notation for applying a Hadamard gate to the n-qubits. The gate labeled $G$ is known as the Grover iteration or Grover operator. Essentially, the Grover operator increases the amplitude of the basis states that are being searched for and decreases the amplitudes of the other states. This is done by rotating the current state vector (of the n-qubits) towards the superposition of all solutions to the search problem ($\ket{\beta}$) as shown in Figure~\ref{fig:gop}.
The Grover operator consists of
\begin{enumerate}
	\item Applying the oracle search function (using the oracle workspace) to the n-qubits. 
	\item Applying Hadamard transformations to all n-qubits
	\item Performing a conditional phase shift
	\item Applying Hadamard transformations to all n-qubits
\end{enumerate}
The initial oracle application reflects the current state of the n-qubits about $\ket{\alpha}$ to move it from its initial state $1$ to state $2$, as shown in Figure~\ref{fig:gop}. The remaining three operations (Hadamard, conditional phase shift, Hadamard) then reflect the state $2$ vector about state $1$. Overall, this achieves a rotation of $\theta$ towards $\ket{\beta}$ (state $3$). Repeating the Grover operator $O(\sqrt{n})$ times rotates the state vector very close to $\ket{\beta}$, the superposition of all solutions. Hence, when the state is measured with reference to the computational basis it reveals a solution to the search problem with extremely high probability\footnote{The exact probability of observing a solution depends upon the number of solutions in the search space.}. This is the essence of Grover's algorithm; an excellent detailed analysis of Grover's algorithm and its recent variations is given in chapter 6 of Nielsen and Chuang's book \cite{Nielsen2000}.
	
	\subsubsection{Deutch's and Scaling Majority-On Problems}
	\label{sec:problemdefn}
	Deutsch's problem is determining whether a given oracle, or black box, function is uniform or balanced, given that the oracle must be either one of these. The uniform property requires an oracle to always return 0 or always return 1; the balanced property requires an oracle to return an equal number of 0s and 1s over all possible inputs. The scalable majority-on problem is an extension of Deutsch's problem where the oracle is an arbitrary boolean function and the problem is to determine whether the majority of the outputs are 1. Quantum algorithms to solve these problems have been evolved by Spector et al. using genetic programming (see Section~\ref{sec:spector}).
	
	\subsection{Further Reading}
	This section presents the major summary references in the field of quantum computing that complement the specific references given in previous subsections. However, most of the references given here have been previously referenced in this paper.
	
	There are currently a large, and ever increasing, number of publications on quantum computing, and consequently there is a lot of duplication and work of varying standards. For issues on quantum computing, Nielsen and Chuang's book \cite{Nielsen2000} appears to be the most comprehensive and well-structured publication, which is considered to be the most significant reference. A very good introductory reference is provided by Landry \cite{Landry2004}, who summarised the major areas in Nielsen and Chuang's book. Another excellent and recommended review reference is Rieffel and Polak's work \cite{RieffelPolak2000}, which contains a more comprehensive and detailed review of quantum computing than Landry's paper \cite{Landry2004}. Furthermore, in contrast to \cite{Landry2004,RieffelPolak2000}, Gram\ss~\cite{Gramss1998} provides a summary of quantum computing from a quantum Turing machine perspective, rather than the quantum circuit model. However, many (although not all) of the topics included in \cite{Gramss1998,Landry2004,RieffelPolak2000} have been covered in this paper.
	
	There are two excellent quantum computing books that also contain detailed information on how to actually simulate quantum algorithms on a classical computer: Williams and Clearwater's book \cite{WilliamsClearwater1998} comes with Mathematica\footnote{Mathematica is a comprehensive mathematical software package, details are available on the their website: http://www.wolfram.com/products/mathematica/index.html.} notebooks that simulate well-known quantum algorithms such as Shor's factoring algorithm, and Hardy and Steeb's book \cite{HardySteeb2001} contains Java and C++ code for some simple quantum simulations such as generating entangled qubits.
	
	The majority of the articles referenced in this quantum computing section are freely available at the Los Alamos preprint server: http://xxx.lanl.gov/archive/quant-ph. This site also provides an excellent place to search for old and new articles within the field of quantum computing.

\section{Introduction to Evolutionary Algorithms}
\label{sec:ea}
	Optimisation problems can be characterised by two sets of parameters: feedback parameters to optimise according to a target solution, and free parameters to modify in order to approach the desired solution \cite{Born1998}. The optimisation algorithm alters the free parameters while controlling the quality of the solution by the feedback parameters; different optimisation techniques perform this search in different ways. When traditional optimisation techniques are used to search vast, complex and unknown spaces there are extremely complex constraints and multimodal problems\footnote{Multimodal problems are problems that arise in cases with a large number of locally optimum solutions.} \cite{HardySteeb2001}. Thus, traditional optimisation techniques are not well-suited for these types of problems, so alternative approaches have been researched. Evolutionary Algorithms (EAs) are one of the alternative methods that have gained significant popularity as general-purpose optimization and search tools \cite{HardySteeb2001}. EAs are probabilistic search algorithms that are heavily based upon Darwinian evolution, as a proxy for the process of species evolution in nature. The central concept of Darwinian evolution is that individuals in a population have heritable characteristics that influence their probability of producing offspring, that is, future generations. EAs have extended this theory slightly to state that characteristics of `better' individuals will increase their likelihood to produce offspring, which is a variation on Darwin's `survival of the fittest' principle \cite{Darwin1859}. The idea behind this is an attempt to converge to the `best' individual, which essentially is the paradigm of search and optimisation.
	
	EAs were initially used as optimisation tools for engineering problems, and were developed independently by several computer science researchers in the 1950s and 1960s \cite{GPT2004}. Since then, the number of applications for EAs has become diverse and has grown at a fast rate \cite{Born1998}; for example, there are EA applications in financial forecasting, predicting protein structure, predicting the primeness of numbers and in developing computer programs. In contrast to the application aspect of EAs, despite much theoretical research there has been modest progress in EA theory over the last 20 years compared with that of neural networks, another biologically motivated form of computation \cite{Belew1997}. Nevertheless, various types of EAs have been developed and the types presented in the following subsection now form the backbone of the EA field \cite{Mitchell1996}. Regardless of what type of EA is used the basic elements are almost identical \cite{BNKF1998}; these common elements are \cite{Leier2004}
	\begin{itemize}
		\item Populations of individuals representing solutions to the problem at-hand, which allows parallel searching.
		\item Ways to manipulate solutions, which can be either
		\begin{itemize}
			\item Mutation (inspired by the biological process of the same name) operators, which implement innovative change
			\item Recombination/Crossover (inspired by the biological process of gamete production and sexual reproduction) operators, which implement conservation of characteristics through rearrangement
		\end{itemize}
		\item A measure for determining the quality of a solution, usually referred to as a fitness function
		\item A method of selection that uses the fitness function to select individuals for the next generation
	\end{itemize}
	
	\subsection{Types of EAs}
	
		\begin{figure}[htb]
\centerline{
   \includegraphics[scale=0.5]{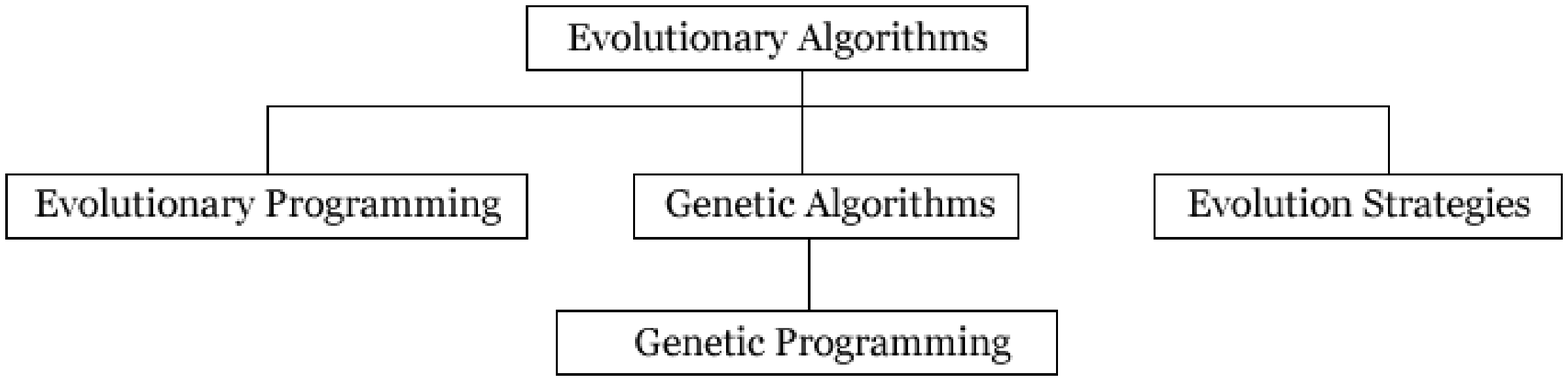}
}
\caption{Relationships between different types of Evolutionary Algorithms}
\label{fig:EAtypes}
\end{figure}	
	
	EAs and Genetic Algorithms (GAs) are terms that are sometimes incorrectly used interchangeably. However, EA is an umbrella term that includes all algorithms that incorporate the idea of Darwinian evolution, and GAs are just one type of EA. The different types of EAs are shown in Figure~\ref{fig:EAtypes}. GAs, invented by Holland in the 1960s \cite{Holland1975}, are the most prominent EAs \cite{HardySteeb2001}, with the other major types of EAs being Evolutionary Programming (EP) and Evolution Strategies (ES). Essentially, EP and ES operate on and change the phenotype (observable properties) of individuals, while GAs operate on the genotype (genetic construction) of individuals \cite{Fogel1995}. To further explain the concept of phenotype, a change in phenotype can be viewed as a change in the behavior, physiology or morphology without altering the genetics. In addition, a further difference between ES and GAs is that ES places an emphasis on mutation \cite{Born1998}, whereas GAs place a higher weighting on recombination. However, having stated the differences between these types of EA, it is important to note that these distinctions are not strict, as overlap does occur \cite{Born1998}. For example, GAs have been applied to real numbered genomes \cite{SurryRad1997}, which overlaps into the ES field. For a more thorough review of EP and ES, which also includes a discussion on GAs, please refer to B{\"a}ck's book \cite{Back1996}.
	
	In 1992, Koza introduced a new type of EA known as Genetic Programming (GP) \cite{Koza1992}; Bornholdt \cite{Born1998} suggests that the reason GP arrived late into the EA field may have been due to the need for greater computing power. GP is a type of GA whereby the search space is reduced to solely include computer programs. GP techniques have been valuable in evolving structures other than computer programs, but the fact that individuals in a GP model are computer programs is the most defining feature of GP \cite{Spector2000}. Nevertheless, GP has extended the idea of genotype manipulation from GAs to include variable length chromosomes; that is, the representation of individuals in a population can be of varying size.
	
	\subsection{GA and GP Algorithm Structure}
		
		\begin{figure}[htb]
\centerline{
   \includegraphics[scale=0.5]{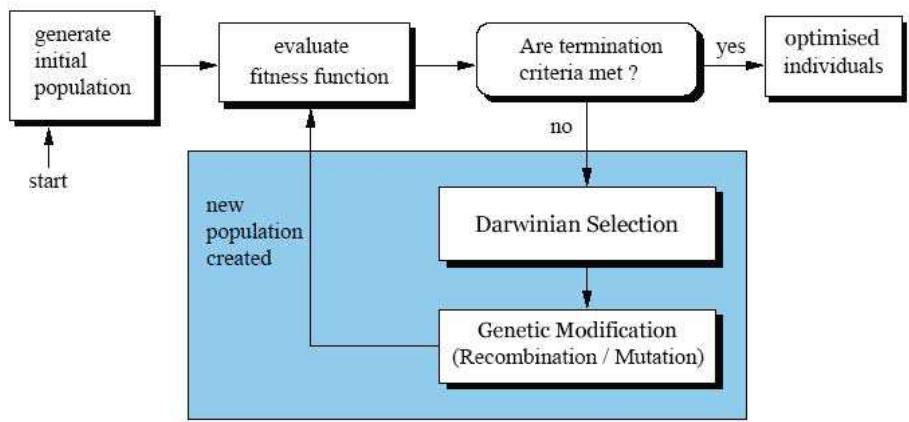}
}
\caption{A general flow diagram of GAs and GP algorithms, modified from Figure 3.1 in Leier's thesis \cite{Leier2004}.}
\label{fig:GAGP}
\end{figure}
		
		The overall structure of GAs and GP algorithms (GPAs) is shown in Figure~\ref{fig:GAGP}. Initially, a starting population is generated; this is usually done randomly so that it results in a diverse population that simulates all naturally occurring populations. However, before an initial population can be created there must be a way to represent individuals. Traditionally, as supported by Holland \cite{Holland1975}, individuals, or solutions, were encoded as binary strings; modern GAs (and consequently GPAs) use tailored encoding to suit the problem being solved \cite{Michal1996}. The number of individuals in the initial population is another parameter that must be set according to the specific problem at hand.
		
		After an initial population is created the evolutionary loop begins. The evolutionary loop consists of the following repeating sequence:
\begin{enumerate}
	\item Evaluate the fitness of each individual in the population. The fitness, or quality, of each individual (x) is measured by a predetermined fitness function ($f(x)$) with a real number codomain, that is, it returns a real number.
	\item Select individuals based on fitness levels to be parents of the next generation, as in Darwinian Selection. The three main selection methods used in GAs and GPAs are
		\begin{enumerate}
			\item \emph{Fitness-Proportional} selection, where individuals are selected using a \emph{roulette wheel} concept in which each individual ($i$) has a sector of size proportional to its fitness ($f_{i}$). Assuming that a lower fitness function corresponds to a superior fitness, this means each individual has a probability of being selected equal to
			\begin{displaymath}
				\frac{1-f_{i}}{\sum_{i}{\left( 1-f_{i}\right)}}
			\end{displaymath}
			\item \emph{Ranking} selection, which simply involves selecting the best x (predefined parameter) number of individuals, according to a predefined function based on their fitness. 
			\item \emph{Tournament} selection, which involves randomly selecting a number (predefined parameter) of individuals to compete in the tournament. The losers of the tournament, based on fitness, are then excluded from selection, and all remaining individuals are selected. This is however, a much weaker form of selection than the ranking selection.
		\end{enumerate}
	\item Generate the next population through genetic modification. The mutation operator is applied to a single individual which results in a random change to its representation (genome) at one or more positions; it is applied with a certain low probability, set as a parameter. The crossover operator is applied to two individuals (parents) and produces a new individual (child) for the next generation, which is a combination of the representations (genomes) of the original individuals. The crossover operator is also applied with a probability set as a parameter, but this probability is usually much higher that the mutation probability. Thus, cross-over can obviously be performed in many ways and depends on the representations of the individuals. Two possible approaches are (1) using the beginning of the representation from one parent and the end from the other parent, or (2) reusing one parent's representation with a middle section exchanged for a section of the other parent. One common classification of different crossover operators can be separated into the following two categories:
		\begin{itemize}
			\item \emph{Fixed-length} crossover, where the children have the same length representations as their parents. Assuming that individuals are a sequence of instructions, an example of fixed-length crossover is taking the first 4 instructions from parent A (length 6) and the last 2 instructions from parent B (length 6), which would produce a child that had 6 instructions, the same as its parents.
			\item \emph{Variable-length} crossover, which simply means that the result of the crossover operation may result in a different length child. For example, taking the first 4 instructions from parent A (length 6) and the last 5 instructions from parent B (length 6) produces a child that had 9 rather than 6 instructions.
		\end{itemize}

\end{enumerate}

		This loop terminates when sufficiently optimised individuals\footnote{The desired level of optimisation is set as a parameter in terms of the fitness function.} have been evolved or other predetermined termination criteria are met, such as the maximum number of generations have been evolved.
		
		\subsubsection{Further Reading}
		For futher information on GAs, Bornholdt \cite{Born1998} presents a succinct review, and Hardy and Steeb \cite{HardySteeb2001} provide a good review of the implementation of GAs, which includes many Java and C++ code examples. In addition, a very comprehensive analysis of GAs from a mathematical standpoint is given by Vose \cite{Vose1999}. More detail on GP in general is covered in Koza's books \cite{Koza1992,Koza1994}.

\section{Evolving Quantum Algorithms}
Genetic Programming (GP) is defined as developing algorithms based on Darwinian evolution to automatically generate computer programs. Thus, as quantum computing is a formal computational model, GP can be used to evolve quantum computing programs (represented as quantum algorithms); this field will be hereafter referred to as `Evolving Quantum Algorithms'. Unlike other applications of GP, there has been relatively little work done in the field of evolving quantum algorithms \cite{Rubinstein2001}.

As was stated in Section~\ref{sec:potent}, it seems that the major reasons for a lack of discovery of quantum algorithms are that they are difficult and non-intuitive to generate, and that only faster than classical quantum algorithms are of interest. Therefore, it would seem appropriate that we look to computers to search for quantum algorithms. However, as the search space of quantum algorithms is vast, complex and reasonably uncharted, using traditional search techniques is problematic (see Section~\ref{sec:ea}). Hence the motivation to use GP to evolve quantum algorithms, with the hope that GP's power to randomly search vast, complex and unknown spaces can discover many superior quantum algorithms.

\subsection{Simulation Limitations}
In order to evolve quantum algorithms the fitness of different individuals (quantum algorithms) must be assessed; the fitness of a solution is assessed by running the algorithm. However, as quantum computers have not yet been built (Section~\ref{sec:QCrealise}), quantum algorithms must be simulated on a classical machine. However, due to quantum parallelism simulating quantum algorithms on a classical computer comes at a potentially exponential simulation cost \cite{Spector1998,Rubinstein2001,LeierBanzhaf2003}. Therefore, as inefficiencies in simulation will only multiply rapidly during the evolutionary process, an efficient quantum computer simulator is a necessity in this field \cite{Leier2004}. Spector \cite{Spector2000} introduces a tested quantum computer simulator termed QGAME, which is based upon the quantum circuit model detailed in Section~\ref{sec:qcircuit}; Sabry \cite{HaskellQModel} outlines an interesting quantum simulation model in the functional language Haskell, and there are also many quantum computer simulators that vary in quality available on the Internet. An example code fragment of the quantum simulator used by Spector et al. \cite{Spector1998} explains a possible representation of gates in the programming language LISP, which is demonstrated below:\\\\
	(HADAMARD 0)  ;; apply Hadamard transformation to qubit 0 \\
	(CNOT 1 2) ;; apply CNOT gate with control (qubit 1) and target (qubit 2)\\\\
In addition, recall that Hardy and Steeb \cite{HardySteeb2000} provide some C++ and Java code for simulating different quantum effects. Nevertheless, due to the simulating inefficiency only small quantum algorithms can be simulated and evolved. In this sense small quantum algorithms means that the following must be restricted in size: number of qubits, number of gates (commonly referred to as the length of a quantum circuit), and the number of iterations of the algorithm (with varying input) needed to determine the fitness of the algorithm. Hence, discovering small, yet scalable, quantum algorithms is the ultimate result from evolving quantum algorithms.

	\subsection{Program Structures}
	\label{sec:structure}
	Within GP, the individuals (computer programs) must have a defined structure so that genetic operators can be applied to them. In most applications of GP, the computer programs consist of basic building blocks, referred to as primitives or genomes, such as constants, operators, problem-specific functions and inputs \cite{Leier2004}. When evolving quantum algorithms the main components of the quantum algorithm are quantum gates. These primitives can then combine to form complete computer programs in different structures; the types of program structures that have been applied in the field of evolving quantum algorithms are \emph{tree}, \emph{linear},  and the hybrid \emph{linear-tree}. The choice of an appropriate program structure appears to be a key ingredient to the success of the GP evolution \cite{Leier2004}.
	
	The original GP program structure, as outlined by Koza \cite{Koza1992,Koza1994}, was based on the standard tree structure. Quantum gates and their arguments are represented as parent nodes and their children respectively. Children nodes can be quantum gates themselves\footnote{The return value of a quantum gate node is a valid index of a qubit it acted upon. For example, the CNOT gate is usually defined to return the index of the control qubit.}, and leaf nodes can be constants, zero-argument functions, qubit indices or other program inputs. This structure can be easily translated into a functional programming language where the functions are the parent nodes and the function's arguments are its children. The program can also be executed with a post-order traversal (left-subtree, right sub-tree, root), which means no further memory is required as function arguments are always locally accessible. It is also important to note that it is trivial to convert a tree-structured quantum program into the quantum circuit model, where the order of gates in the circuit model will be defined by the post-order traversal\footnote{There is also a trivial mapping of a quantum circuit to a tree-structured program, based on establishing a sequence of gates, that is further described in Leier's thesis \cite{Leier2004}.}.
	
	Although Koza's tree structure provides elegant adaptations to different program sizes and shapes \cite{Langdon1999} and allows for easy expression in any functional programming language, there is no guarantee it is the most appropriate representation for all problems \cite{Spector1999}. An alternative to the tree structure, is a linear structure that unsurprisingly consists of a linear sequence of instructions. Spector et al. \cite{Spector1999} outline reasons that suggest tree-based program structures offer no advantages over linear program structures, but have additional complexity. They state that quantum algorithms are linear, so a linear program structure appears to be more appropriate than a tree structure. Operators or functions in the linear based model then get their arguments from external memory. The linear-tree structure is a combination of these two structures, where there is a linear structure and branching function at each node of the tree structure. During execution of the tree structure, the linear structure of each node visited is executed. The tree structure execution begins by visiting the root node and executes only one path to a leaf node: the next node visited is determined by the branching function at the current node. In quantum computing, partial measurements of the entire quantum state are appropriate branching functions \cite{Leier2004}; therefore, the next node depends on the result of the partial measurement. An example branching function is if the measurement of qubit $n$ yields $\ket{0}$ or $\ket{1}$, then the left or right subtree will be visited next respectively. 
	
	It should also be noted that there is a \emph{graph}, and consequently a \emph{linear-graph}, GP program structure that has been applied to areas other than quantum algorithms. There are many types of graph structures: the well-known PADO graph structure is presented by Teller and Veloso \cite{Teller}. It seems likely that the reason this structure has not been used to evolve quantum algorithms is that the complexity cost is greater than that of other structures, but this approach does not seem to represent quantum effects better than other structures.
	
	\subsection{Previous Studies}
	In addition to the papers cited, Leier's PhD thesis \cite{Leier2004} is a major source for this section and provides more details on the specific parameters of each study than are given in this section. A new paper by Giraldi, Portugal and Thess \cite{GPT2004} also contains, among other things, reviews of Rubinstein's \cite{Rubinstein2001} and Yabuki and Iba's \cite{YabukiIba} work; however, Leier's thesis \cite{Leier2004} appears to be the best reference.

		\subsubsection{Williams and Gray}
		The book by Williams and Gray \cite{WilliamsGray1997} is the pioneering work that suggests the use of genetic programming to evolve quantum algorithms. However, the goal of their study was to search more efficiently for alternative quantum circuits to a known quantum algorithm  than conducting an exhaustive search. The difference from subsequent studies is that, in this case the overall unitary gate that represents the quantum algorithm was already known. This is still useful as it could be used to search for (hopefully faster) alternative quantum algorithms that solve the same problem \cite{Rubinstein2001}. In this case, the fitness function would include a penalty for circuits that were similar to the known quantum algorithm.
		
		Using a population size of 100, the GP algorithm implemented in this study successfully found a quantum teleportation algorithm (see Section~\ref{sec:teleport} for details on quantum teleportation), where the send portion was as efficient, and the receive portion more efficient, than the known algorithm; the GP algorithm used:
		\begin{itemize}
			\item A linear program structure
			\item An approximate universal-gate set consisting of the CNOT gate and two other one-qubit gates
			\item A three-tuple representation: \{quantum gate, [parameters\footnote{Only discrete parameters were allowed.}]\footnote{[ ] indicates optionality.}, set of qubits acted upon\}
			\item A fitness function based on the evolved circuits similarity to the known solution
			\item A ranking selection scheme
			\item Mutation and Crossover operators that act on quantum gate(s)
		\end{itemize}
	
	\subsubsection{Yabuki and Iba}
		Yabuki and Iba \cite{YabukiIba} developed a GA model\footnote{It is technically not a GP model due to the fixed length representation of individuals.} that was specifically tailored to evolve the quantum teleportation algorithm as did Williams and Gray \cite{WilliamsGray1997}. Using a larger population size of 5000, the GA algorithm implemented in this study successfully found a quantum teleportation algorithm (see Section~\ref{sec:teleport} for details on quantum teleportation) that has at least 3 fewer gates than any non-evolved quantum teleportation algorithm; the GA algorithm used:
		\begin{itemize}
			\item A linear program structure
			\item The same gate set as Williams and Gray \cite{WilliamsGray1997} (chosen as the problem was the same)
			\item A unique \emph{fixed-length} string representation that was specifically tailored to the production of two entangled particles, which is essential in quantum teleportation
			\item A problem specific fitness function based on the error of the evolved algorithm's output
			\item A fitness-proportional selection scheme
			\item Mutation and fixed-length crossover operators
		\end{itemize}
		
		Yabuki and Iba, like Williams and Gray, have approached evolving quantum algorithms from an optimisation, rather than a search, standpoint. They have focused on using a tailored evolutionary algorithm to find a more efficient quantum algorithm for a given existing quantum algorithm. Furthermore, along with Williams and Gray's \cite{WilliamsGray1997} results, it has been shown that there is potential in this area, particularly for the quantum teleportation algorithm.
	
		\subsubsection{Spector et al.} 
		\label{sec:spector}
		Spector has recently published the first book \cite{Spector2004} about using GP to automatically evolve quantum algorithms, which makes reference to all of his previous work with a more detailed introduction into the field. Spector et al. \cite{Spector1998,SpectorBB2000,SpectorBBS1999,Spector1999} have conducted extensive research on evolving quantum algorithms. Three different GP models were outlined and applied to various problems in \cite{Spector1999}. The common elements of all three models are the
\begin{itemize}
	\item Standardised fitness function (detailed in \cite{Spector1998}) that takes into account three components:
		\begin{enumerate}
			\item \emph{Hits}: total number of fitness cases minus the number of fitness cases for which the quantum circuit yields the correct answer with less than 0.48 probability
			\item \emph{Correctness}: $\frac{\sum_{i=1}^{numCases}{max\left(0,error_{i}-0.48\right)}}{max\left(hits,1\right)}$. This formula deliberately ignores any circuits with errors less than 0.48, so that the focus is on producing correct quantum algorithms, not on improving the probability of correctness of already correct quantum algorithms\footnote{Here Spector et al. are using the term correct to mean correct greater than or equal to 48\% of the time.}
			\item \emph{Efficiency}: number of quantum gates / 100,000, which makes sure that this component will always be less than 1.
		\end{enumerate}
		Spector et al. recommend that these components be combined lexicographically, whereby quantum circuits are only compared on Correctness if their Hits are the same, and similarly only compared on Efficiency if the Hits and Correctness are the same.
	\item Tournament selection method with a tournament size of five individuals
\end{itemize}
		
		The first GP model defined in \cite{Spector1999} uses a standard tree program structure, and is in fact a summary of the results presented in Spector et al.'s initial paper \cite{Spector1998}. This GP model was applied to two oracle problems, namely Deutsch's problem and the scaling majority-on problem (defined in Section~\ref{sec:problemdefn}). Using a population size of $10,000$, this tree structure GP algorithm was used to evolve a better-than-classical quantum algorithm for Deutsch's problem, but did not evolve a quantum algorithm to solve the majority-on problem better than any classical algorithm. Along with the common elements shown above, this tree structure GP algorithm also used
		\begin{itemize}
			\item A set of gate building functions as well as iteration structures and arithmetic operators
			\item Functions in prefix notation, analogous to functional languages such as LISP in which the quantum algorithms were simulated
			\item Mutation operators, and two types of crossover operators: a variable length operator design for the tree structure and a reproduction operator to produce a child exactly the same as a parent.
		\end{itemize}
		
		The other two GP models use two types of linear program structures. As stated in Section~\ref{sec:structure}, a linear structure stores arguments in external memory. The first type of linear structure, called the stack-based linear GP, uses a global stack for temporary data storage. This GP model successfully evolved a quantum algorithm to solve the four-item database search problem faster than any classical algorithm; moreover, the evolved quantum algorithm is as efficient as, and almost identical to, Grover's search algorithm. Besides the program structure, the differences between this GP model and the tree structure model discussed above are that the
	\begin{itemize}
		\item Various crossover operators are designed to operate on linear, rather than tree, structures; they are also fixed-length operators
		\item Iteration structures are stack-based, rather than tree-based
		\item Gate building functions do not return values on to the stack
		\item Arithmetic functions take their arguments from, and return their result to, the stack
		\item A $no$--$op$ operator is part of the function set
	\end{itemize}
	
	The primary role of including iteration structures is to produce scalable quantum algorithms, however, some non-scalable quantum algorithms are of interest \cite{Spector1999}. The second type of linear GP model is tailored to evolving these non-scalable quantum algorithms as there is no iteration structure. Furthermore, Spector et al. \cite{Spector1999} suggest that there is no major reason for quantum gates to share parameter values, which consequently means there is no reason for data storage. Thus, in contrast to the stack-based linear structure, the second type of linear structure implemented, called the stack-less linear GP, only contains the gate building functions and has no external memory for temporary data storage. This GP structure was used to evolve a quantum algorithm to solve the And-Or Query problem faster than any previously known classical or quantum algorithm. The And-Or Query problem is to determine whether the boolean function $(f(0)\vee f(1))\wedge(f(2)\vee f(3))$ is true or false for a two-bit boolean function $f$. 
				
		Spector et al. \cite{SpectorBB2000,SpectorBBS1999} then presented a modification of their stack-less linear structure GP. The changes to the GP model include
\begin{itemize}
	\item Using a steady-state GP. All previous GP models applied to quantum algorithms have been generational GP models; the difference is simply that steady-state GP models do not have clearly defined generations. The remainder of the evolutionary process can be considered to be very similar
	\item Supporting the use of variable-length representations and variable-length crossover operators
	\item A four component fitness function, in which the components are again combined lexicographically; the components are the
		\begin{enumerate}
			\item \emph{Misses} component, which is equivalent to the \emph{Hits} component used in \cite{Spector1999}
			\item \emph{Expected-Queries} component that considers the number of oracle calls (defined in \cite{SpectorBBS1999}), which is a tailored modification for evolving quantum algorithms to solve oracle problems
			\item \emph{Max-Error} component, which is the maximum probability of getting a wrong answer in any fitness case, and is similar to the \emph{Correctness} component used in \cite{Spector1999}
			\item \emph{Num-Gates} component, which is equivalent to the \emph{Efficiency} component used in \cite{Spector1999}
		\end{enumerate}
	\item Including a one-qubit measurement gate so that partial measurements can be made, which are key to several known quantum algorithms such as Shor's factoring algorithm
	\item The added ability to distribute the evolutionary process across multiple workstations to decrease execution time
\end{itemize}
		This modified GP model was again applied to the And-Or Query problem, with better results that the initial stack-less linear structure model. From these improving results, Spector et al. deduced that they were successfully improving their GP model to evolve quantum algorithms, and that the stack-less linear structure is probably the best structure that has been developed for evolving quantum algorithms. Furthermore, the successful research by Spector et al. summarised above has shown that there definitely is a degree of potential in evolving quantum algorithms.
	
		\subsubsection{Rubinstein}
		Using a population size of 5000, Rubinstein \cite{Rubinstein2001} used his generational GP algorithm to successfully discover the most efficient known quantum algorithms to produce two to five maximally entangled qubits\footnote{Two maximally entangled qubits is in fact the first Bell state as described in Section~\ref{sec:entangle}.}, that is, qubits of the form $\frac{1}{\sqrt{2}}\left(\ket{0\ldots0}+\ket{1\ldots1}\right)$. The GP algorithm used
		\begin{itemize}
			\item A linear program structure (with no external storage)
			\item An unspecified set of quantum gates that include CNOT, Hadamard and importantly an `Observe' gate that can measure one or many qubits, which is known to be a vital technique in several known quantum algorithms, such as Shor's factoring algorithm
			\item A modification of the three-tuple representation, where the quantum gate and its parameters and qubit operands are encoded into a bit string, which is the standard representation for an individual in GPs
			\item A fitness function based solely on the error of the evolved quantum system. The fitness was calculated using the following formula, where $i$ is an input case of $k$ total cases, $j$ is a basis state in a quantum system of $n$ qubits, and $o$ and $d$ are the observed and desired amplitudes respectively.
			\begin{displaymath}
			error = \sum_{i}^{k-1}{\sum_{j}^{2n-1}{\left(o_{ij}-d_{ij}\right)}}
			\end{displaymath}
			The error obtained from this function is then divided by the highest error of any individual evolved to obtain a stardardised fitness that lies in the range between 0 (optimal) and 1.
			\item A fitness-proportional selection scheme
			\item Mutation operators (with low probability), and crossover operators that act upon all parts of the bit string representation, that is, gates, parameters and qubits. Gate cross-over is variable-length, but parameter and qubit crossover are fixed-length, as most quantum gates act upon a fixed number of qubits with a fixed number of parameters
		\end{itemize}
		
		From the quantum algorithms produced for the maximum entanglement problem with two to five qubits, deductions were made about an arbitrary sized maximum entanglement production circuit. Thus, this study shows that there is potential for GP to produce small, yet scalable, quantum circuits that can be converted into large scale quantum circuits.		
		
	\subsubsection{Lukac et al.}
	Lucak and Perkowski \cite{LukacP2002} present a general GA for evolving quantum circuits. Their algorithm used:
	\begin{itemize}
			\item An encoding system where quantum circuits are represented as an array of strings of quantum gates. Each element of the array represented a specific point in time (after the previous element and before the next element) in the quantum algorithm when gates could act upon qubits. The string of ordered gates at each element corresponded to the order of qubits, such that the the first gate acted upon the first qubit and so on. This system had no extra parameters to identify which qubits a gate acted upon, thus it was only possible with the introduction of a no operation gate. Their encoding system also deliberately allowed for parallel evalution of individuals to potentially decrease fitness evaluation time.
			\item A large gate set comprising various one, two and three qubit gates, including a one-qubit `wire' gate that performs `no operation'. The reason for the diverse gate set is Lukac and Perkowski's focus on evolving arbitrary quantum algorithms.
			\item A fitness function considering the correctness of the quantum algorithm. This fitness function is similar to those in previous papers \cite{Rubinstein2001,WilliamsGray1997}, but in this case a fitness level of 1 corresponds to maximum (not minimum) fitness.
			\item Roulette wheel and stocastic universal sampling selection schemes. Stocastic universal sampling is less biased towards selecting 'fit' individuals, thus a more random selection scheme (which is detailed more in \cite{Gold1989}).
			\item Mutation and crossover operators.
			\item Population sizes of 50 and 100 individuals. The population was deliberately kept small to avoid long fitness evaluation times.
	\end{itemize}
	
	Lucak and Perkowski used their algorithm to evolve various quantum circuits, averaging each experiment over 20 runs. First, they evolved single gate circuits (whereby the target gate was included in the gate set for the experiment) to test their algorithms convergence. All gates were successfully evolved, but for larger number of input-qubits (gate size) longer evolution time was required. Interestingly, mutation with a probability greater than 0.4 was found to decrease both real-time evolution and the number of generations needed to successfully evolve the target gate. Their second experiment was to evolve three quantum circuits consisting of more than one gate: namely, the quantum teleportation algorithm previous evolved by Williams and Gray \cite{WilliamsGray1997} and three and four maximally entagled qubits as previously evolved by Rubinstein \cite{Rubinstein2001}. In all three cases, the desired quantum circuit was evolved in similar or less time than previously published. It was also noted that while the number of generations required to evolve quantum circuits increases exponentially as the number of qubits increase, the real-time evolution increases at a much slower rate. Higher probability of mutation was again found to decrease the number of generations and real-time for successful evolution. 
	
	While not discovering any new or further optimised quantum algorithms, Lucak and Perkowski have established benchmarks for the evolution of various small quantum algorithms, ranging from 1 to 4 qubits. Furthmore, for the three composite quantum algorithms ($>1~qubit$) evolved, their GA has performed equally or more efficiently (in terms of time of evolution) than previous studies. Lukac et al. \cite{LukacP2003} have furthered their research by investigating implementing their GA on a quantum computer. The specific encoding used in this GA lends itself to be computer in parallel, thus it should be implementable on quantum technologies, such as nuclear magnetic resonance (NMR).
		
		\subsubsection{Leier}
		Leier, in his recent PhD thesis \cite{Leier2004}, presented two GP models for evolving quantum algorithms, one linear and one linear-tree structure. Both models are very similar to the successful stack-less linear structure GP model developed by Spector et al. \cite{SpectorBBS1999,SpectorBB2000}. The major difference is the four component fitness function of Leier's model, in which the components are combined through different weightings, rather than lexicographically. The actual components are also slightly different, but have all been previously been mentioned by Spector et al.; they are: \emph{misses}, \emph{max-error}, \emph{correctness} and \emph{num-gates}. Further differences in the linear-tree model, caused by its structural difference, are that branching functions are partial measurements and the crossover operators that act on the linear and tree substructures are included. The inherent inclusion of partial measurements and the added flexibility of the linear-tree structure was the motivation for creating a GP model that was not strictly linear. 
		
		Both GP models were applied to the Deutsch-Jozsa (D-J) and 1-SAT problems\footnote{Note that the result of the linear-tree models applied to 1-SAT is also contained in a paper by Leier and Banzhaf \cite{LeierBanzhaf2003}.}. The Deutsch-Jozsa problem is essentially a scalable version of Deutsch's problem; 1-SAT also has a known better-than-classical quantum algorithm solution (Hogg's algorithm). The linear-tree GP was able to find a quantum algorithm essentially the same as the known algorithms for both the 1-SAT and D-J problem, although some evolutionary runs did not produce a solution to the D-J problem. The interesting finding from this study was that intermediate partial measurements had no noticeable positive effect; similarly, the added flexibility of the linear-tree structure did not add a benefit over the strict linear structure which also evolved solutions to both problems. However, it is probable that these findings can not be generalised past these small problems that have relatively simple quantum solutions. Nevertheless, this study was the first to show that linear-tree structure GP models can be used to evolve quantum circuits.
		
		Leier also made some other interesting observations; the most important of these are listed below:
		\begin{itemize}
			\item Using the linear structured GP model, Leier showed that there was an increase in the efficiency of the evolutionary process when using a pre-evolved, rather than a random, initial population for both the D-J and 1-SAT problem. This pre-evolution involved feeding evolved individuals from a smaller problem instance into the initial population.
			\item According to Leier's research crossover is not as important as commonly thought. This finding conflicts with the traditional GP approach, where crossover is performed with much greater probability than mutation with the idea of multiplying and distributing better solutions over the population.
			\item Even though Leier made a point of emphasising that his GP models were not designed to produce scalable quantum algorithms, scalability was achieved: the algorithms evolved using Leier's models, which apply to n-qubits, can be easily manually scaled to apply to (n+1)-qubits. However, this is in fact not surprising as the known quantum algorithms for both application problems are scalable.
		\end{itemize}	
	
	\subsubsection{Ding et al.}
	Ding et al. \cite{DingJinYang2006} recently presented a new framework for evolving quantum circuits that is designed for both quantum algorithm discovery and optimisation. This framework uses a Hybrid Quantum-Inspired Evolutionary Algorithm (HQEA), which was motivated by GP and detailed in Yang's Masters thesis \cite{Yang2006}. Ding et al.'s approach used:
	\begin{itemize}
			\item A fixed length numerical encoding of quantum circuits, compared with encoding in symbols as done in previous works. Encoding the quantum circuits with numerical values was necessary to take advantage of the HQEA algorithm.
			\item A gate set comprising the Hadamard, CNOT, Phase and $\pi/8$ gates\footnote{Within the scope of this paper, Phase and $\pi/8$ gates can be thought of as reflection (about a basis state) and rotation gates respectively.}. Their approach is not limited to these gates, but gates in their quantum algorithms are confined to one-qubit and adjacent two-qubit gates. For example, the control qubit of the CNOT gate must be adjacent to the qubit undergoing the NOT operation.
			\item A fitness function that considers both the cost and correctness of quantum algorithms (as also done in Reid's Masters thesis \cite{Reid2005}). The fitness function used, where lower fitness is better, was:
				\begin{displaymath}
					fitness=reward\times(actualcost - satcost)+punish\times(1-correctness)
				\end{displaymath}
			The $satcost$ represents a satisfactory algorithms cost, whereby if it is set high or low then the evolution is inclined towards discovery or optimisation respectively. Algorithm cost was calculated with one-qubit gates, two-qubit gates, and the wire costing 1, 2 and 0 respectively. However, using this same framework algorithm cost could be more accurately computed in terms of the monetary costs using different quantum technologies.
			\item A fitness-proportional selection scheme
			\item Mutation and crossover operators.
	\end{itemize}
		
	Ding et al. tested their approach on evolving 2 and 3 entangled qubits, as well as the controlled-phase gate, which confirmed that lower $satcosts$	resulted in more optimised quantum algorithms. However, more generations of evolution were required to evolve optimised algorithms. Further experiments were conducted on evolving 2 entagled qubits as research into the appropriate values for the reward-punish factors. This revealed that a large punish:reward ratio ($\geq 5~for~satcost=6)$ is required to evolve correct quantum algorithms, and larger punish is required for larger $satcost$. In addition, Ding et al. present a faster method for matrix multiplication with tensor product, to increase the efficiency of evaluating individuals. This faster method can also be adapted to other evolutionary algorithms. Overall, this new approach has shown promise in evolving and optimising small quantum algorithms, but it has not yet been used to discover a previously undiscovered quantum algorithm.		
	
	\subsection{Further Reading}
	The papers mentioned in the previous section provide further reading: Spector et al.'s work that was published in a book \cite{Spector1999} is recommended as the first paper to read due to its understandability. Spector's book \cite{Spector2004} and Leier's excellent PhD thesis \cite{Leier2004} are recommended as the main points of call on the topic of evolving quantum algorithms. Both of these references also contain an introduction to both the field of quantum computing and evolutionary algorithms, which is similar to, but more detailed than, the one given in this paper. Spector's paper \cite{Spector2000} is also recommended as an excellent short paper that extends the idea of evolving quantum algorithms using GP to evolving arbitrary computational processes.
	
		\subsubsection{Related Applications of GP}
		Genetic programming has also been applied to some other closely-related applications. Quantum program evolution based on density matrices \cite{Stadel2004} is a new area that is showing promise. Spector et al. \cite{Spector1998} also suggested that there was potential in researching how GP algorithms would be executed on quantum computers, and whether a significant speed-up on fitness evaluation is possible using quantum parallelism. This idea has been reviewed by Giraldi, Portugal and Thess \cite{GPT2004} in 2004, and Udrescu et al. \cite{UdrescuPV2006} have presented new research into implementing GA algorithms on quantum computers. Another of Spector et al.'s suggestions is to evolve hybrid quantum-classical algorithms, whereby the quantum algorithm has classical pre and post data processors \cite{Spector1998}. Another somewhat related paper \cite{RJBSS2004} looked into a case of using GAs to evolve the hardware of quantum computers: a set of pulse sequences (which can be thought of as rotations) is evolved for a given quantum logic gate, implemented by Nuclear Magnetic Resonance (NMR). The significance of this is that the shorter and more robust the pulse sequence the more efficient the implementation of a quantum gate or algorithm; furthermore, the evolved set of pulse sequences was superior to any previously known set.
	
\section{Conclusion}
Genetic algorithms and programming have been successfully used to analyse and optimise known quantum algorithms. Previous studies have also shown that genetic programming can evolve new quantum algorithms, albeit only small quantum algorithms. Another possible outcome of evolving quantum algorithms is that a new useful idea, such as a meaningful sequence of gates, which will change the way future quantum algorithms are developed manually, may be discovered; however as yet, this has not occurred. Nevertheless, manual quantum algorithm generation has had more success than evolving quantum algorithms \cite{Leier2004}, although no further breakthroughs have been made manually since Shor's and Grover's algorithms. Furthermore, the fact that quantum computers do not yet exist is a huge limitation on the research field of evolving quantum algorithms, as quantum simulation on a classical machine has an exponential order overhead. This means that only small quantum algorithms can be evolved, which are few in number and generally have little practical application. Thus, Leier and Banzhaf \cite{Leier2004,LeierBanzhaf2003} speculate that this field will have a brighter future when quantum computers exist. However, there are still ideas that have not been tested as the field of evolving quantum algorithms is relatively new and has not yet been comprehensively researched \cite{Leier2004}.

\bibliographystyle{acm}
\bibliography{AllQuant}

\begin{thebibliography}{10}

\bibitem{Back1996}
{\sc B{\"a}ck, T.}
\newblock {\em Evolutionary Algorithms in Theory and Practice: Evolution
  Strategies, Evolutionary Programming, Genetic Algorithms}.
\newblock Oxford Univeristy Press, 1996.

\bibitem{BNKF1998}
{\sc Banzhaf, W., Nordin, P., Keller, R., and Francone, F.}
\newblock {\em Genetic Programming -An Introduction}.
\newblock dpunkt Heidelberg and Morgan Kaufmann Publishers, San Francisco,
  1998.

\bibitem{Barenco1995}
{\sc Barenco, A.}
\newblock A universal two-bit gate for quantum computation.
\newblock In {\em Proceedings of the Royal Society of London A 449\/} (1995),
  pp.~679--683.

\bibitem{Lots1995}
{\sc Barenco, A., Bennett, C.~H., Cleve, R., DiVincenzo, D.~P., Margolus, N.,
  Shor, P., Sleator, T., Smolin, J., and Weinfurter, H.}
\newblock Elementary gates for quantum computation.
\newblock {\em Physical Review A 52}, 5 (1995), 3457--3467.

\bibitem{Belew1997}
{\sc Belew, R.~K., and Vose, M.~D.}, Eds.
\newblock {\em Foundations of Genetic Algorithms 4\/} (1997), Morgan Kaufmann.

\bibitem{Ben1980}
{\sc Benioff, P.}
\newblock The computer as a physical system: A microscopic quantum mechanical
  hamiltonian model of computers as represented by turing machines.
\newblock {\em Journal of Statistical Physics 22}, 5 (1980), 563--591.

\bibitem{Bennett1973}
{\sc Bennett, C.~H.}
\newblock Logical reversibility of computation.
\newblock {\em {IBM} Journal of Research and Development 17\/} (1973),
  525--532.

\bibitem{Bennett1997}
{\sc Bennett, C.~H., Bernstein, E., Brassard, G., and Vazirani, U.~V.}
\newblock Strengths and weaknesses of quantum computing.
\newblock {\em Society for Industrial and Applied Mathematics Journal on
  Computing 26\/} (1994), 1510--1523.

\bibitem{BV1993}
{\sc Bernstein, E., and Vazirani, U.~V.}
\newblock Quantum complexity theory.
\newblock In {\em Proceedings of the 25th Annual {ACM} Symposium on Theory of
  Computation\/} (1993), pp.~11--20.

\bibitem{Born1998}
{\sc Bornholdt, S.}
\newblock Genetic algorithms.
\newblock In {\em Non-Standard Computation}, T.~Gram{\ss}, S.~Bornholdt,
  M.~Gro{\ss}, M.~Mitchell, and T.~Pellizzari, Eds. {WILEY-VCH}, Weinheim,
  Germany, 1998, pp.~141--178.

\bibitem{Bouw1997}
{\sc Bouwmeester, D., Pan, J.-W., Mattle, K., Eibi, M., Weinfurther, H., and
  Zeilinger, A.}
\newblock Experimental quantum teleportation.
\newblock {\em Nature 390\/} (1997), 575--579.

\bibitem{Braun1995}
{\sc Braunstein, S.~L.}
\newblock Quantum computation: a tutorial.
\newblock Tech. rep., Department of Computer Science, York University, 1995.

\bibitem{Darwin1859}
{\sc Darwin, C.}
\newblock {\em On the origin of species by means of natural selection or the
  preservation of favoured races in the struggle for life}.
\newblock Murray, London, 1859.

\bibitem{Deutsch1985}
{\sc Deutsch, D.}
\newblock Quantum theory, the church-turing principle and the universal quantum
  computer 400.
\newblock In {\em Proceedings of the Royal Society of London A\/} (1985),
  pp.~97--117.

\bibitem{Deutsch1989}
{\sc Deutsch, D.}
\newblock Quantum computational networks.
\newblock In {\em Proceedings of the Royal Society of London A 425\/} (1989),
  pp.~73--90.

\bibitem{DeutschJozsa1992}
{\sc Deutsch, D., and Jozsa, R.}
\newblock Rapid solution of problems by quantum computation.
\newblock In {\em Proceedings of the royal society of london series A\/}
  (1992), vol.~A439, pp.~553--558.

\bibitem{DingJinYang2006}
{\sc Ding, S., Jin, Z., and Yang, Q.}
\newblock Evolving quantum oracles with hybrid quantum-inspired evolutionary
  algorithm.
\newblock {\em ArXiv Quantum Physics e-prints\/} (October 2006).

\bibitem{Dirac1958}
{\sc Dirac, P.}
\newblock {\em The Principles of Quantum Mechanics}, fourth~ed.
\newblock Oxford University Press, 1958.

\bibitem{Feynman1982}
{\sc Feynman, R.}
\newblock Simulating physics with computers.
\newblock {\em International Journal of Theoretical Physics 21\/} (1982),
  467--488.

\bibitem{Feynman1959}
{\sc Feynman, R.~P.}
\newblock There's plenty of room at the bottom: An invitation to enter a new
  field of physics.
\newblock Speech at the annual meeting of the {\emph{American Physical
  Society}}, December 1959.
\newblock It is available online at http://www.zyvex.com/nanotech/feynman.html.

\bibitem{Feynman1961}
{\sc Feynman, R.~P., and Gilbert, D.~H.}, Eds.
\newblock {\em Miniaturization}.
\newblock Reinhold, New York, 1961, pp.~282--295.

\bibitem{Fogel1995}
{\sc Fogel, D.~B.}
\newblock Phenotypes, genotypes, and operators in evolutionary computation.
\newblock In {\em Computational Intelligence: Theory and Applications, 5th
  Fuzzy Days\/} (Berlin, 1995), Springer-Verlag, pp.~337--342.

\bibitem{ForneyGuha2005}
{\sc {Forney, Jr}, G.~D., and Guha, S.}
\newblock Simple rate-1/3 convolutional and tail-biting quantum
  error-correcting codes.
\newblock submitted to 2005 IEEE International Symposium on Information Theory,
  2005.

\bibitem{FortnowRogers1999}
{\sc Fortnow, L., and Rogers, J.}
\newblock Complexity limitations on quantum computation.
\newblock {\em Journal of Computer and System Sciences 59}, 2 (1999), 240--252.
\newblock Special issue for selected papers from the 13th IEEE Conference on
  Computational Complexity.

\bibitem{GPT2004}
{\sc Giraldi, G.~A., Portugal, R., and Thess, R.~N.}
\newblock Genetic algorithms and quantum computation.
\newblock {\em CoRR cs.NE/0403003\/} (2004).

\bibitem{Gold1989}
{\sc Goldberg, D.~E.}
\newblock {\em Genetic Algorithms in Search, Optimisation, and Machine
  Learning}.
\newblock Addison Wesley, 1989.

\bibitem{Gramss1998}
{\sc Gram{\ss}, T.}
\newblock The theory of quantum computation: An introduction.
\newblock In {\em Non-Standard Computation}, T.~Gram{\ss}, S.~Bornholdt,
  M.~Gro{\ss}, M.~Mitchell, and T.~Pellizzari, Eds. {WILEY-VCH}, Weinheim,
  Germany, 1998, pp.~141--178.

\bibitem{Gross1998}
{\sc Gro{\ss}, M.}
\newblock Molecular computing.
\newblock In {\em Non-Standard Computation}, T.~Gram{\ss}, S.~Bornholdt,
  M.~Gro{\ss}, M.~Mitchell, and T.~Pellizzari, Eds. {WILEY-VCH}, Weinheim,
  Germany, 1998, pp.~15--58.

\bibitem{Grover1996}
{\sc Grover, L.~K.}
\newblock A fast quantum mechanical algorithm for database search.
\newblock In {\em Proceedings of the 18th annual ACM symposium on the history
  of computing\/} (Philadelphia, Pennsylvania, May 1996), pp.~212--219.

\bibitem{Gudder1999}
{\sc Gudder, S.}
\newblock Quantum automata: An overview.
\newblock {\em International Journal of Theoretical Physics 28}, 9 (1999),
  2261--2282.

\bibitem{HardySteeb2000}
{\sc Hardy, Y., and Steeb, W.-H.}
\newblock Entangled quantum states and a {C++} implementation.
\newblock {\em International Journal of Modern Physics C 11\/} (2000), 69--77.

\bibitem{HardySteeb2001}
{\sc Hardy, Y., and Steeb, W.-H.}
\newblock {\em Classical and Quantum Computing, with C++ and Java Simulations}.
\newblock {Birkh\"auser} Verlag, Berlin, Germany, 2001.

\bibitem{Holland1975}
{\sc Holland, J.~H.}
\newblock {\em Adaptation in Natural and Artificial Systems}.
\newblock MIT Press, Cambridge, 1975.

\bibitem{Hunger1974}
{\sc Hungerford, T.~A.}
\newblock {\em Algebra}.
\newblock Springer Verlag, New York, 1974.

\bibitem{Koza1994}
{\sc Koza, J.~R.}
\newblock {\em Genetic Programming II}.
\newblock MIT Press, Cambridge, 1992.

\bibitem{Koza1992}
{\sc Koza, J.~R.}
\newblock {\em Genetic programming: on the programming of computers by means of
  natural selection}.
\newblock MIT Press, Cambridge, 1992.

\bibitem{Landry2004}
{\sc Landry, O.}
\newblock Introduction to quantum computing.
\newblock From Physics Department of McGill University, April 2004.

\bibitem{Langdon1999}
{\sc Langdon, W.~B., Soule, T., Poli, R., and Foster, J.~A.}
\newblock The evolution of size and shape.
\newblock In Spector et~al. \cite{AdvanceGP}, pp.~163--190.

\bibitem{Leier2004}
{\sc Leier, A.}
\newblock {\em Evolution of quantum algorithms using genetic programming}.
\newblock PhD thesis, University of Dortmund: Department of Computer Science,
  2004.

\bibitem{LeierBanzhaf2003}
{\sc Leier, A., and Banzhaf, W.}
\newblock Evolving {H}ogg's quantum algorithm using linear-tree {GP}.
\newblock In {\em Genetic and Evolutionary Computation -- GECCO-2003\/}
  (Chicago, 12-16 July 2003), E.~Cant{\'u}-Paz, J.~A. Foster, K.~Deb, D.~Davis,
  R.~Roy, U.-M. O'Reilly, H.-G. Beyer, R.~Standish, G.~Kendall, S.~Wilson,
  M.~Harman, J.~Wegener, D.~Dasgupta, M.~A. Potter, A.~C. Schultz, K.~Dowsland,
  N.~Jonoska, and J.~Miller, Eds., vol.~2723 of {\em LNCS}, Springer-Verlag,
  pp.~390--400.

\bibitem{Lenstra1993}
{\sc Lenstra, A.~K., and {Lenstra, Jr}, H.~W.}
\newblock {\em The Development of the Number Field Sieve}.
\newblock Lecture Notes in Mathematics 1554. Springer Verlag, 1993.

\bibitem{Loss1998}
{\sc Loss, D., and DiVincenzo, D.}
\newblock Quantum computation with quantum dots.
\newblock {\em Physical Review A 57\/} (1998), 120--126.

\bibitem{LukacP2002}
{\sc Lukac, M., and Perkowski, M.}
\newblock Evolving quantum circuits using genetic algorithms.
\newblock In {\em Proceedings of the 2002 NASA/DOD Conference on Evolvable
  Hardware\/} (2002), pp.~177--185.

\bibitem{LukacP2003}
{\sc Lukac, M., Perkowski, M., Kerntopf, P., Pivtoraiko, M., Folgheraiter, M.,
  Lee, D., Kim, H., Hwuangbo, W., wook Kim, J., and Choi, Y.~W.}
\newblock A hierarchical approach to computer aided design of quantum circuits.
\newblock In {\em Proceedings of the 6th International Symposium on
  Representations and Methodology of Future Computing Technology\/} (2003),
  pp.~201--209.

\bibitem{Michal1996}
{\sc Michalewicz, Z.}
\newblock {\em Genetic algorithms + Data structures = Evolution Programs},
  3rd~ed.
\newblock Springer - Verlag, New York, 1996.

\bibitem{Mitchell1996}
{\sc Mitchell, M.}
\newblock {\em An introduction to genetic algorithms}.
\newblock MIT Press, 1996.

\bibitem{Nielsen2000}
{\sc Nielsen, M.~A., and Chuang, I.~L.}
\newblock {\em Quantum computation and Quantum Information}.
\newblock Cambridge University Press, Cambridge, 2000.

\bibitem{Pellizzari1998}
{\sc Pellizzari, T.}
\newblock Quantum computers: first steps towards a realization.
\newblock In {\em Non-Standard Computation}, T.~Gram{\ss}, S.~Bornholdt,
  M.~Gro{\ss}, M.~Mitchell, and T.~Pellizzari, Eds. {WILEY-VCH}, Weinheim,
  Germany, 1998, pp.~141--178.

\bibitem{Purple}
{\sc Pittenger, A.~O.}
\newblock {\em An Introduction to Quantum Computing Algorithms}.
\newblock Birk{h\"a}user, Boston, 2000.

\bibitem{PRS1998}
{\sc P\u{a}un, G., Rozenberg, G., and Salomaa, A.}
\newblock {\em DNA Computing: New Computing Paradigms}.
\newblock EATCS Series. Springer-Verlag, 1998, pp.~1--6,43--74.

\bibitem{DNA1998}
{\sc P\u{a}un, G., Rozenberg, G., and Salomaa, A.}
\newblock {\em DNA Computing: New Computing Paradigms}.
\newblock EATCS Series. Springer-Verlag, 1998.

\bibitem{Reid2005}
{\sc Reid, T.}
\newblock On the evolutionary design of quantum circuits.
\newblock Master's thesis, Waterloo, Ontario, Canada, 2005.

\bibitem{RJBSS2004}
{\sc Rethinam, M.~J., Javali, A.~K., Behrman, E.~C., Steck, J.~E., and Skinner,
  S.~R.}
\newblock A genetic algorithm for finding pulse sequences for nmr quantum
  computing.
\newblock {\em submitted to Physical Review A\/} (April 2004).

\bibitem{RieffelPolak2000}
{\sc Rieffel, E., and Polak, W.}
\newblock An introduction to quantum computing for non-physicists.
\newblock {\em {ACM} Computing Surveys\/} (2000).

\bibitem{Rubinstein2001}
{\sc Rubinstein, B. I.~P.}
\newblock Evolving quantum circuits using genetic programming.
\newblock In {\em Proceedings of the 2001 IEEE Congress on Evolutionary
  Computation (CEC2001)\/} (2001), IEEE Press, pp.~114--121.

\bibitem{HaskellQModel}
{\sc Sabry, A.}
\newblock Modeling quantum computing in {Haskell}.
\newblock In {\em Haskell '03: Proceedings of the ACM SIGPLAN workshop on
  Haskell\/} (New York, NY, USA, 2003), ACM Press, pp.~39--49.

\bibitem{Shor1994}
{\sc Shor, P.~W.}
\newblock Algorithms for quantum computation: Discrete log and factoring.
\newblock In {\em Proceedings of the 35th Annual Symposium on Foundations of
  Computer Science\/} (November 1994), Institute of Electrical and Electronic
  Engineers Computer Society Press, pp.~124--134.

\bibitem{Shor1995}
{\sc Shor, P.~W.}
\newblock Scheme for reducing decoherence in quantum computer memory.
\newblock {\em Physical Review A 52\/} (1995), 2493--2496.

\bibitem{Shor2000}
{\sc Shor, P.~W.}
\newblock Introduction to quantum algorithms.
\newblock Notes for talk given for the short course at the January 2000
  American Math Society meeting, 2000.

\bibitem{Simon1994}
{\sc Simon, D.}
\newblock On the power of quantum computation.
\newblock In {\em Proceedings of the 35th annual IEEE symposium on the
  foundations of computer science (FOCS)\/} (Santa Fee, USA, November 1994),
  IEEE Computer Society Press, pp.~116--123.

\bibitem{Spector2000}
{\sc Spector, L.}
\newblock The evolution of arbitrary computational processes.
\newblock {\em IEEE Intelligent Systems\/} (May/June 2000), 80--83.

\bibitem{Spector2004}
{\sc Spector, L.}
\newblock {\em Automatic Quantum Computer Programming: A Genetic Programming
  Approach}.
\newblock Genetic Programming Series. Kluwer Academic Publishers, 2004.

\bibitem{Spector1998}
{\sc Spector, L., Barnum, H., and Bernstein, H.~J.}
\newblock Genetic programming for quantum computers.
\newblock In {\em Genetic Programming 1998: Proceedings of the Thurd Annual
  Conference\/} (1998), J.~R. Koza, W.~Banzhaf, K.~Chellapilla, K.~Deb,
  M.~Dorigo, D.~B. Fogel, M.~H. Garzon, D.~E. Goldberg, H.~Iba, and R.~L.
  Riolo, Eds., pp.~365--374.

\bibitem{SpectorBB2000}
{\sc Spector, L., Barnum, H., and Bernstein, H.~J.}
\newblock Quantum circuits for or and and of ors.
\newblock {\em Journal of Physics A: Mathematical and General 33\/} (2000),
  8047--8057.

\bibitem{SpectorBBS1999}
{\sc Spector, L., Barnum, H., Bernstein, H.~J., and Swamy, N.}
\newblock Finding a better-than-classical quantum {AND/OR} algorithm using
  genetic programming.
\newblock In {\em Proceedings of the 1999 Congress on Evolutionary
  Computation\/} (1999).

\bibitem{Spector1999}
{\sc Spector, L., Barnum, H., Bernstein, H.~J., and Swamy, N.}
\newblock Quantum computing applications of genetic programming.
\newblock In Spector et~al. \cite{AdvanceGP}, pp.~135--160.

\bibitem{AdvanceGP}
{\sc Spector, L., Langdon, W.~B., O-Reilly, U.-M., and Angeline, P.~J.}, Eds.
\newblock {\em Advances in Genetic Programming: Volume 3}.
\newblock The MIT Press, 1999.

\bibitem{Stadel2004}
{\sc Stadelhofer, R.}
\newblock Solving the parity problem on a mixed state quantum computer.
\newblock Tech. rep., University of Dortmund, 2004.
\newblock Available upon request.

\bibitem{SClark2005}
{\sc Stepney, S., and Clark, J.~A.}
\newblock Evolving quantum programs and protocols.
\newblock In {\em Handbook of Theoretical and Computational Nanotechnology},
  M.~Rieth and W.~Schommers, Eds. American Scientific Publishers, 2005.
\newblock Overview available at
  {http://www-users.cs.york.ac.uk/\%7Esusan/bib/ss/nonstd/gqprev05.htm}.

\bibitem{SurryRad1997}
{\sc Surry, P., and Radcliffe, N.}
\newblock A formalism for real-parameter evolutionary algorithms and directed
  recombination.
\newblock In Belew and Vose \cite{Belew1997}.

\bibitem{Teller}
{\sc Teller, A., and Veloso, M.}
\newblock {PADO}: {A} new learning architecture for object recognition.
\newblock In {\em Symbolic Visual Learning}, K.~Ikeuchi and M.~Veloso, Eds.
  Oxford University Press, 1996, pp.~81--116.

\bibitem{UdrescuPV2006}
{\sc Udrescu, M., Prodan, L., and Vl\u{a}du\c{t}iu, M.}
\newblock Implementing quantum genetic algorithms: a solution based on grover's
  algorithm.
\newblock In {\em Proceedings of the Third Conference on Computing frontiers\/}
  (New York, 2006), {ACM} Press, pp.~71--82.

\bibitem{vanDam1996}
{\sc van Dam, W.}
\newblock A universal quantum cellular automaton.
\newblock In {\em Proceedings of PhysComp96\/} (1996), New England Complex
  Systems Institute, pp.~323--331.

\bibitem{Tonder2004}
{\sc {van Tonder}, A.}
\newblock A lambda calculus for quantum computation.
\newblock {\em SIAM Journal on Computing 33}, 5 (2004), 1109--1135.

\bibitem{Nature2001}
{\sc Vandersypen, L. M.~K., Steffen, M., Breyta, G., Yannoni, C.~S., Sherwood,
  M.~H., and Chuang, I.~L.}
\newblock Experimental realization of shor's quantum factoring algorithm using
  nuclear magnetic resonance.
\newblock {\em Nature 414\/} (December 2001), 883--887.

\bibitem{Vose1999}
{\sc Vose, M.~D.}
\newblock {\em The simple genetic algorithm: foundations and theory}.
\newblock The MIT Press, 1999.

\bibitem{Watrous1998}
{\sc Watrous, J.}
\newblock Relationships between quantum and classical space-bounded complexity
  classes.
\newblock In {\em 13th Annual {IEEE} Conference on Computational Complexity\/}
  (June 1998), pp.~210--227.

\bibitem{WilliamsGray1997}
{\sc Williams, C., and Gray, A.}
\newblock {\em Automated Design of Quantum Circuits}.
\newblock Springer, New York, 1997.

\bibitem{WilliamsClearwater1998}
{\sc Williams, C.~P., and Clearwater, S.~H.}
\newblock {\em Explorations in Quantum Computing}.
\newblock Springer Verlag, 1998.

\bibitem{Woot1982}
{\sc Wooters, W.~K., and Zurek, W.~H.}
\newblock A single quantum cannot be cloned.
\newblock {\em Nature 299\/} (1982), 802--803.

\bibitem{YabukiIba}
{\sc Yabuki, T., and Iba, H.}
\newblock Genetic algorithms for quantum circuit design, evolving a simpler
  teleportation circuit.
\newblock In {\em GECCO-00: Procedings of the Genetic and Evolutionary
  Computation Conference\/} (San Francisco, July 2000), Morgan Kauffman
  Publishers, pp.~421--425.

\bibitem{Yang2006}
{\sc Yang, Q.}
\newblock The research of a hybrid quantum-inspired evolutionary algorithm.
\newblock Master's thesis, Wuhan University, China, 2006.

\bibitem{Yao1993}
{\sc Yao, A.}
\newblock Quantum circuit complexity.
\newblock In {\em Proceedings of the 34th Annual Symposium on the Foundations
  of Computer Science\/} (Los Alamitos, USA, 1993), {IEEE} Computer Society
  Press, pp.~352--361.

\bibitem{Zurek1991}
{\sc Zurek, W.~H.}
\newblock Decoherence and the transition from quantum to classical.
\newblock {\em Physics Today 44\/} (1991), 36--44.

\end{thebibliography}

\end{document}